\documentclass[12pt,preprint]{aastex}
\shorttitle{$\alpha$-Enhanced Simple Stellar Populations Models}
\shortauthors{H.-c. Lee et al.}

\begin{document}

\title{Comparison of Alpha-Element Enhanced Simple Stellar Population 
Models with Milky Way Globular Clusters}

\author{Hyun-chul Lee}
\affil{Department of Physics and Astronomy, Washington State University, 
Pullman, WA 99164-2814; and Department of Physics and Geology, 
The University of Texas - Pan American, Edinburg, TX 78539}
\email{hclee@wsu.edu}
\author{Guy Worthey}
\affil{Department of Physics and Astronomy, Washington State University, 
Pullman, WA 99164-2814}
\author{Aaron Dotter}
\affil{Department of Physics and Astronomy, University of Victoria, 
Victoria, BC, V8W 3P6, Canada}

\begin{abstract}

We present simple stellar population (SSP) models with scaled-solar
and $\alpha$-element enhanced abundances. The SSP models
are based on the Dartmouth Stellar Evolution Database, our library 
of synthetic stellar spectra, and a detailed systematic variation 
of horizontal-branch (HB) morphology with age and metallicity.  
In order to test the relative 
importance of a variety of SSP model ingredients, 
we compare our SSP models with integrated spectra of 41 Milky Way
Globular Clusters (MWGCs) from Schiavon et al. (2005).
Using the Mg $b$ and Ca4227 indices, we confirm that Mg and Ca are
enhanced by about +0.4 and +0.2 dex, respectively, in agreement with 
results from high resolution spectra of individual stars in MWGCs.  
Balmer lines, particularly H$\gamma$ and H$\delta$, of MWGCs are 
reproduced by our $\alpha$-enhanced SSP models not only because of 
the combination of isochrone and spectral effects but also because of
our reasonable HB treatment.  Moreover, it is shown that the Mg abundance 
significantly influences Balmer and iron line indices.  Finally, 
the investigation of power-law initial mass function (IMF) variations 
suggests that an IMF much shallower than Salpeter is unrealistic 
because the Balmer lines are too strong on the metal-poor side 
to be compatible with observations.  

\end{abstract}

\keywords{stars: abundances --- stars: evolution --- 
stars: horizontal-branch --- globular clusters: general}

\section{Introduction}

Understanding the star formation and chemical enrichment histories 
of galaxies from their integrated spectrophotometries is one of the 
overarching goals of modern astronomy and astrophysics.  Over the 
last two decades, we have learned that massive elliptical galaxies 
have enhanced $\alpha$-element abundances, such as Mg, compared to the 
Sun (Worthey et al. 1994; Worthey 1998; Lee \& Worthey 2005).  
According to $\Lambda$CDM cosmology, it is usually suggested that 
star formation was most intense at the deep potentials inhabited
by massive ellipticals.  Their intense star formation skewed the 
chemical abundance pattern toward the yields from massive 
core-collapse supernovae. 

Central to this line of investigation, the Lick/IDS spectral indices 
have been widely used to derive the mean age and metallicity 
of stellar systems such as star clusters and galaxies.   
It is therefore essential to validate the simple stellar population 
(SSP; single age and single metallicity) models by 
comparing conclusions derived from theoretical integrated Lick 
spectral indices with results obtained from resolved stellar 
population studies.  The most suitable nearby sample is arguably 
the Milky Way globular clusters (MWGCs).  They are generally old, span 
a broad range of metallicity, and show some $\alpha$-element 
enhancement.  

There have been many efforts in the literature to understand 
the integrated spectroscopic properties of MWGCs 
(e.g., Burstein et al. 1984; Gibson et al. 1999; Vazdekis et al. 2001; 
Lee, Yoon, \& Lee 2000; Schiavon et al. 2002; Thomas et al. 2003; 
Lee \& Worthey 2005).  Although there is mounting evidence that the 
MWGCs are perhaps not an ideal representatives of SSPs 
(e.g., Piotto 2009; Yi 2009), they are still very useful targets 
for the purpose of testing SSP models.  

In order to compute the theoretical integrated Lick/IDS spectral indices 
from SSP models for the broad range of age and metallicity, 
three components are generally used.  They are (1) stellar isochrones 
and/or evolutionary tracks, (2) stellar spectra, 
and (3) empirically derived Lick spectral index fitting functions.  
The fitting functions of \citet{wor94b}\footnote{
\citet{wo97} for H$\delta_{A}$, H$\gamma_{A}$, H$\delta_{F}$, 
H$\gamma_{F}$.  There are different sets of fitting functions 
in the literature (e.g., Schiavon 2007).} have been built 
upon the Lick stellar library that basically follows the 
galactic chemical enrichment histories throughout the broad 
metallicity coverage.  In a nutshell, stars on the metal-poor 
side ([Fe/H] $\leq$ $-$ 1.0) are generally $\alpha$-element enhanced, 
[$\alpha$/Fe] $\sim$ +0.4 dex 
and near solar metallicity they are solar-scaled, 
i.e., [$\alpha$/Fe] = 0.0.  

Stellar evolutionary models and high resolution synthetic spectral 
libraries have been updated frequently.  Some recent examples 
of $\alpha$-element enhanced stellar models are Kim et al. (2002), 
Pietrinferni et al. (2006), VandenBerg et al. (2006), and 
Dotter et al. (2007b).  In this work, we present scaled-solar 
and $\alpha$-element enhanced SSP models for a broad range of 
age and metallicity that employ the Dotter et al. (2007b) stellar 
evolution models and the Lee et al. (2009) high resolution 
synthetic stellar spectra.  A detailed systematic variation 
of horizontal-branch (HB) morphology with age and metallicity is 
incorporated in order to faithfully represent the MWGCs.  
Lee et al. (2000) investigated the effects from HB stars with 
a detailed variation of HB morphologies as a function of age and 
metalicity.  Their input ingredients were, however, the amalgamation 
of the Revised Yale isochrones with Yi et al. (1997) HB evolutionary 
tracks.  In this work, our HB morphology treatment is based on 
the recent study by Dotter (2008).  This work differs from 
recent efforts by Coelho et al. (2007) and Schiavon 
(2007) in that we cover a broader range of metallicity and age 
and adopt different isochrones, stellar spectra, and reasonable 
treatment of HB morphology\footnote{
Coelho et al. (2007)'s [Fe/H] range is only from $-$0.5 to +0.2 and 
that of Schiavon (2007) is from $-$1.3 to +0.3.  Schiavon (2007) 
focused only for the blue side of the spectra, 
so he did not cover Fe5406 index.}.  

Milky Way globular clusters are known for their different degree of 
$\alpha$-element enhancement from scaled-solar chemical 
abundances (e.g., Pritzl, Venn, \& Irwin 2005).  
Thus it is important to self-consistently include these abundances
patterns at every stage of SSP model construction.  
We examine our new models with the large, homogeneous dataset of MWGC 
integrated spectra compiled by Schiavon et al. (2005; hereafter S05).  
Earlier observational efforts in this category include 
Burstein et al. (1984), Cohen et al. (1998), and Puzia et al. (2002).
However, the S05 dataset is the larget and most representative of the
MWGC population.

In the following section, we describe our new models in detail. 
The results from our models are depicted in $\S$ 3.  
Our models are compared with Lick indices measured from the S05 data
in $\S$ 4.  Section 5 summarizes and concludes this work.

\section{Models}

The present models are direct descendents of an evolutionary
population synthesis code that was developed to study the stellar
populations of globular clusters and galaxies 
(\citealt{lee00}; \citealt{lee02}; \citealt{lee04}; \citealt{lw05}; 
\citealt{lee09}).  In this work, we take advantage of the 
recent emergence of isochrones and HB evolutionary tracks with 
$\alpha$-element enhancement at fixed [Fe/H] from \citep{dot07b} 
accompanied with Lee et al. (2009) high resolution synthetic spectra.  
We describe how we generate 
reasonable HB morphologies in section 2.2.  The standard \citet{sal55} 
initial mass function (IMF) was employed for calculating the relative 
number of stars along the isochrones.  The variation of IMF is, however, 
investigated and depicted in section 2.3.  The investigated 
ages are 8, 11, and 14 Gyr 
and the metallicities cover $-2.5 \leq$ [Fe/H] $\leq +0.0$ with 
[$\alpha$/Fe] = 0.0 and +0.4. 

We defer the detailed studies on the effects of multiple populations 
(Piotto 2009), He-enhancement (Yi 2009), CNONa chemical inhomogeneity 
(Pietrinferni et al. 2009), blue stragglers (Cenarro et al. 2008), and 
binary stars in GCs to future work.

\subsection{$\alpha$-Element Enhancement}

In this study, we are illuminating the differences between scaled-solar
and +0.4 dex $\alpha$-element enhancement at fixed [Fe/H] in the SSP models 
by varying (1) isochrones alone, (2) synthetic spectra alone, and (3) both 
isochrones and spectra.

\subsubsection{$\alpha$-enhanced isochrones}

The stellar evolution models that we employ were described by 
\citet{dot07b} and we refer the reader to that paper for complete 
details.  We have used their stellar evolution models 
at [Fe/H] = $-$2.5, $-$2.0, $-$1.5, $-$1.0, $-$0.5, and 0 with 
[$\alpha$/Fe] = 0.0 and +0.4 and ages from 8 to 14 Gyr.  These are 
complemented by a self-consistent set of He-burning tracks that extend 
from the zero-age HB (ZAHB) to the onset of thermal pulsations on the 
asymptotic giant branch (AGB).  In the models, $\alpha$-enhancement refers 
to enhancements in the $\alpha$-capture elements O, Ne, Mg, Si, S, Ca, and 
Ti by the same amount as specified in [$\alpha$/Fe].  The models assume that 
the initial He mass fraction follows $Y$ = 0.245 + 1.54$Z$, where $Y$ = 0.245 
is the primordial He abundance from big bang nucleosynthesis 
(Spergel et al. 2003) and the required initial He abundance of the 
calibrated solar model is $Y$ = 0.274.  

It must be emphasized that our $\alpha$-enhanced stellar models 
in this work are set at fixed [Fe/H], rather than fixed metal mass fraction $Z$, 
so the $\alpha$-elements add additional metals and make the isochrones 
cooler in general.  
In this setting, the temperature differences are more significant 
on the metal-rich side than on the metal-poor as 
illustrated in Figure 1.  The scaled-solar and +0.4 dex $\alpha$-enhanced 
stellar models of Dotter et al. (2007b) are contrasted in the H-R diagrams 
at 8, 11, and 14 Gyr in Figure 1.  The left panel shows the 
$\alpha$-element enhancement at [Fe/H] = $-$2.0, while the right panel 
displays that at [Fe/H] = 0.0.

\subsubsection{$\alpha$-enhanced synthetic spectra}

Investigations into the effects of individual elements on 
stellar spectra, with the intent of applying the results to galaxy 
spectra, include such works as \citet{trip95} and \citet{korn05}, 
who gauged the effects of ten individual elements on the Lick system 
of 25 pseudo-equivalent width indices 
introduced by \citet{wor94b} and \citet{wo97}.  In order to 
investigate the effects of $\alpha$-element enhancement 
on the Lick/IDS indices at solar and super-solar metallicity, 
Lee \& Worthey (2005) employed the updated response functions by 
\citet{hou02}, which expanded the earlier work of Tripicco \& Bell 
(1995) regarding the sensitivity of each Lick spectral indices as 
the abundances of individual chemical elements are varied.  

The newly compiled synthetic spectra that we employ in this study 
were described in \citet{lee09} and we refer the reader to that paper 
for details.  As noted in Lee et al. (2009), the synthetic spectra 
are not very accurate in absolute predictions \citep[c.f.][]{korn05,swb05} 
hence we employ a differential approach in which the fitting functions of
\citet{wor94b} and \citet{wo97} are used as the zero point, and delta-index 
information as a function of element ratio is incorporated via 
measuring the synthetic spectral  library.  This procedure is similar
to that of previous investigations (e.g.,
\citealt{t00a,t00b,ps02,tmb03,lw05,sch07}) but more sophisticated since
an entire grid of delta-index information was used as opposed to 2 
or 3 synthetic stars at solar abundance as in previous works.

The Lick stellar library generally follows the galactic 
chemical enrichment, i.e., [$\alpha$/Fe] = +0.4 for [Fe/H]
$\leq$ $-$1.0, [$\alpha$/Fe] = +0.2 for [Fe/H] = $-$0.5, 
and [$\alpha$/Fe] = 0.0 for [Fe/H] = 0.0 (e.g., \citealt{red06}).  
In order to {\em compensate for this intrinsic abundance trend rooted in 
the stellar library}, we have employed the synthetic spectra 
of [$\alpha$/Fe] = $-$0.4 for [Fe/H] $\leq$ $-$1.0, that of 
[$\alpha$/Fe] = $-$0.2 for [Fe/H] = $-$0.5, and that of 
[$\alpha$/Fe] = 0.0 for [Fe/H] = 0.0 for our {\em solar-scaled} SSP 
models.  Then for our {\em 0.4 dex $\alpha$-enhanced} SSP models, 
we have used the synthetic spectra of [$\alpha$/Fe] = 0.0 
for [Fe/H] $\leq$ $-$1.0, that of [$\alpha$/Fe] = +0.2 
for [Fe/H] = $-$0.5, and that of [$\alpha$/Fe] = +0.4 
for [Fe/H] = 0.0.

\subsection{Horizontal-Branch Morphologies}

Following our earlier work (Lee et al. 2000, 2002, 2004; 
Lee \& Worthey 2005), we fully account for the detailed 
systematic variation of HB morphology with age and metallicity.  
In Lee et al. (2000) and Lee et el. (2002), we basically 
followed the Reimers massloss formula which led to a variation 
in massloss with age and metallicity.  In this work, 
we have instead employed the Dotter (2008) mass loss scheme 
to represent the MWGC HB morphologies.  We have adopted a massloss 
of 0.16 $M_{\odot}$ for [Fe/H] = $-$1.5, $-$1.0, $-$0.5, and 0.0;
0.08 $M_{\odot}$ for [Fe/H] = $-$2.0; and 0.015 $M_{\odot}$ 
for [Fe/H] = $-$2.5 from Figure 3 of Dotter (2008).  
0.02 $M_{\odot}$ is used for the mass dispersion.  
It is important to recognize that Dotter (2008) was focused on 
a set of similarly old age GCs in that paper so that there was 
no claim of an age dependence on mass loss.  
Nevertheless, this trend should be fine so long as it is applied to old 
GCs.  Examples of HB morphologies in the H-R diagram are 
shown in Figure 1 at [Fe/H] = $-$2.0 (left panel) and 0.0 (right panel) 
for 11 Gyr.  The small squares are HB stars for the solar-scaled models, 
while the crosses are the +0.4 dex $\alpha$-enhanced models.  
Similar to the isochrones, the $\alpha$-enhanced HB stars at 
fixed [Fe/H] are also generally cooler than their solar-scaled 
counterparts. 

The number of HB stars are estimated from the number for stars 
that evolved off the RGB using the Salpeter (1955) IMF.  
We find that our $R$-parameter values (i.e., the number of HB 
stars over the number of RGB stars at the zero-age HB stars) 
are similar to Salaris et al. (2004).  Besides the number of HB 
stars, there are several parameters that can vary the HB 
morphologies at given age and metallicity such as different mass 
dispersion, different He, different mass-loss.  Observationally, 
indeed, there are variations of HB morphologies at given [Fe/H] 
within the GC population.  We assign, however, a single HB 
morphology at a given age and metallicity 
by delegating a mass dispersion, and a mass-loss as 
described above.  The purpose of this study is not to investigate 
the detailed match between our SSP models and the observations of 
every single GC, but rather to examine the overall match for the 
entire range of metallicity.

\subsection{IMF Variations}

For the calculation of the relative number of stars along the isochrones, 
we have adopted the standard \citet{sal55} power-law IMF, where 
the number of stars ($d N$) in mass interval ($dm$) is described by 
\begin{equation}
d N \propto m^{-\chi} dm,
\end{equation}
with $\chi$ = 2.35 in this study.  However, we have also investigated the 
variation of IMF as depicted in the below.  The above simple power-law IMF 
with $\chi$ = 2.35 has been varied with exponents of 
$\chi$ = 3.35 (dwarf star-dominant = bottom-heavy IMF) and 
1.35 (giant star-dominant = top-heavy IMF).  We find that the 
$\chi$ = 3.35 case decreases the RGB number density down to 20\%, 
and the $\chi$ = 1.35 case increases the RGB number density up to 
375\%, relative to that of the standard Salpeter IMF with $\chi$ = 2.35.  
As the HB star number density scales with the RGB stars, 
it is seen in the bottom right panels of Figures 2 $-$ 12 that the HB 
effect becomes significantly more important for the top-heavy IMF.

\section{Differential Model Comparisons} 

Having depicted the theoretical aspects of generating the scaled-solar and 
$\alpha$-enhanced Lick spectral indices in $\S$ 2, we now present 
the results of our SSP models.  The effects due to 
$\alpha$-element enhancement in the isochrones and spectra, HB stars, 
and IMF variation on several Lick indices are illuminated in Figures 
2 $-$ 12. The figures are organized as follows: 
In the top left panel, the differences 
between scaled-solar and $\alpha$-enhanced models 
(both isochrones and spectra are enhanced) without HB stars 
(aa-woHB $-$ ss-woHB) are displayed.  The top right panel shows 
the effect of HB stars on the $\alpha$-enhanced models 
(aa-wHB $-$ aa-woHB).  
The middle left panels display isochrone effects alone 
without HB stars (as-woHB $-$ ss-woHB), while the middle right panels 
show spectral effects alone without HB stars (sa-woHB $-$ ss-woHB).  
Thus, the top left panels illustrate the combination of the middle 
panels.  Table 1 lists the {\em spectral effects} of 0.4 dex enhancements of 
individual $\alpha$-elements as well as that of whole $\alpha$-element 
at 11 Gyr, without the consideration of HB stars, at [Fe/H] = $-$1.0 
and 0.0 for our investigated Lick indices.  

In Figures 2 - 5, 7, and 9 - 12, we have compared our model results with 
Coelho et al. (2007) at [Fe/H] = $-$0.5 and 0.0 using their Tables 15 and 17.  
Their SSP models are for 12 Gyr and they are depicted with diamonds.  
Because of differences in isochrones 
and synthetic spectra that are used in this study and Coelho et al. (2007), 
we find some differences.  Regarding the isochrones, there are mixing length 
differences between two studies, $\alpha_{ML}$ = 1.6 for Coelho et al. (2007) 
and 1.938 for Dotter et al. (2007b).  According to Figure 1 of Yi (2003), 
larger $\alpha_{ML}$ makes RGB temperature warmer.  There are also differences 
among the treatment of overshooting and diffusion.  Regarding the synthetic 
spectra, Coelho et al. (2007) are mainly based on ATLAS model atmosphere 
as shown their Figure 5, but Lee et al. (2009) are mainly based on MARCS 
model atmosphere for the similar range of temperature and luminosity.  
These dissimilar model inputs collectively make some different 
model outputs as shown in Figures 2 - 5, 7, and 9 - 12.  It would be 
interesting to see how Coelho et al. models are compared with our models 
and Milky Way globular clusters when their models are extended toward 
the metal-poor regime. 

The bottom panels of Figures 2 $-$ 12, 
depict the variation of IMF effect (the exponent, $\chi$ of the 
power-law IMF in equation (1) between 1.35 and 3.35) without 
(bottom left panels) and with HB stars (bottom right panels) at 11 Gyr 
for our 0.4 dex $\alpha$-enhanced models.

\subsection{Mg $b$ and Ca4227}

Mg $b$:  

The top left panel of Figure 2 shows that Mg $b$ becomes significantly 
stronger with $\alpha$-element enhancement as a function of metallicity.  
It is demonstrated in the middle panels of Figure 2 that both 
$\alpha$-enhanced isochrones and spectra make Mg $b$ stronger as a 
function of [Fe/H], but the latter effects dominates.  The top panels 
show that HB stars do not influence Mg $b$ very much compared 
to $\alpha$-element enhancement and, in fact, make Mg $b$ slightly weaker.
The bottom left panel of Figure 2 shows that Mg $b$ 
increases about 0.2 \AA\ with a bottom-heavy IMF and decreases about 
0.1 \AA\ with a top-heavy IMF; the bottom right panels shows that the IMF
effect is slightly reduced if HB stars are included.  
It is understood that low-mass main sequence stars make Mg $b$ strong.

Ca4227:  

Similar to Mg $b$, 
it is shown in the middle panels of Figure 3 that both 
$\alpha$-enhanced isochrones and spectra make Ca4227 stronger 
as a function of [Fe/H] but, as with Mg $b$, the latter effect dominates.  
Also the top panels display that HB stars do not affect Ca4227 
much compared to abundance effects in the models, as with Mg $b$.  
The bottom panels show that IMF effect is not very significant 
though the top-heavy IMF becomes noticeable because of the 
enhanced effect of HB stars.   

Table 1 indicates that Mg $b$ and Ca4227 are predominantly affected 
by Mg and Ca among $\alpha$ elements, respectively.  At 
[Fe/H] = 0.0, Ca4227 is also sensitive to oxygen though only about 25\% 
of the effect is due to calcium.

\subsection{Balmer Line Indices}

H$\beta$: 

It can be seen in the top right panel of Figure 4 that H$\beta$ 
increases as much as 0.7 \AA\ on the metal-poor side ([Fe/H] $<$ $-$1.0) 
for old stellar populations due to blue HB stars.  Also, it is noted 
from the middle panels of Figure 4 that the $\alpha$-element effects 
come mostly from the cooler isochrones, which weaken the temperature 
sensitive H$\beta$.  The spectral effects are almost negligible, 
except at solar metallicity.  According to Table 1, it is Mg that 
dilutes the H$\beta$ at [Fe/H] = 0.0.  
The bottom left panel displays that 
H$\beta$ increases with a bottom-heavy IMF on the metal-rich side 
([Fe/H] $\geq$ $-$1.0), but the bottom right panel shows that 
the effect of HB stars become considerably more important with a 
top-heavy IMF, particularly on the metal-poor side ([Fe/H] $<$ $-$1.0).

H$\gamma$, H$\delta$: 

Similar to H$\beta$, the top right panels of Figures 5 and 6 show 
that H$\gamma_A$ and H$\gamma_F$ increase as much as 2.4 \AA\ and 
1.4 \AA, respectively on the metal-poor side ([Fe/H] $<$ $-$1.0) 
for old stellar populations due to blue HB stars.  Also, it is 
noted from the middle panels of Figures 5 and 6 that the isochrone 
and spectral effects due to $\alpha$-element are opposite.  
According to Table 1, H$\gamma$ becomes stronger with Mg-, O-, 
and Si-enhancement, while it becomes weaker with Ti-enhancement.  
The spectral effects are much less significant for H$\gamma_F$ 
than for H$\gamma_A$ as noted in Lee et al. (2009).  

Similar to H$\gamma_A$ and H$\gamma_F$, the top right panels of 
Figures 7 and 8 show that H$\delta_A$ and H$\delta_F$ increase 
as much as 1.8 \AA\ and 1.2 \AA, respectively on the metal-poor 
side ([Fe/H] $<$ $-$1.0) for old stellar populations due to blue HB stars.  
Also, it is seen from the middle panels of Figures 7 and 8 that the 
isochrone and spectral effects due to $\alpha$-enhancement are opposite, 
similar to H$\gamma$.  According to Table 1, H$\delta_A$ becomes stronger 
with Mg-, Si-, Ca-, and O-enhancement in decreasing order of importance, 
while becomes weaker with Ti-enhancement.  On the contrary, H$\delta_F$ 
becomes stronger with Si-, Ca-, and Mg-enhancement in decreasing order of 
importance, while becomes weaker with Ti-enhancement.  It 
is noted from Table 1 that the effect of O-enhancement is minimal 
in H$\delta_F$ 
among H$\gamma$ and H$\delta$ indices.  Furthermore, the effect of 
Mg-enhancement on H$\delta_F$ considerably attenuates compared to 
that on H$\delta_A$.  

The IMF effect in the bottom panels of Figures 5 $-$ 8 is similar to 
what we denoted in Figure 4 for H$\beta$.  The top-heavy IMF makes 
both H$\gamma$ and H$\delta$ considerably stronger on the metal-poor 
side ([Fe/H] $<$ $-$1.0) mostly because of 
the enhanced effect of HB stars.  Without the consideration of HB 
stars, the bottom left panel of Figure 8 displays that the IMF effect 
on H$\delta_F$ is minimal among Balmer lines.

\subsection{Iron Line Indices}

Fe5270, Fe5335, Fe5406, Fe4383:\footnote{It becomes clear in 
Lee et al. (2009) that Fe4531 and Fe5015 are Ti-sensitive indices.}

The top left panel of Figure 9 displays that Fe5270 becomes 
stronger with $\alpha$-element enhancement except at the solar 
metallicity.  The middle panels of Figure 9 illustrate that,  
while the cooler $\alpha$-enhanced isochrones at fixed [Fe/H] 
make Fe5270 stronger at all metallicity, the spectral effect makes 
Fe5270 weaker on the metal-rich side.  
According to Table 1, Ca-enhancement makes Fe5270 stronger 
on the metal-poor side but it is Mg-enhancement which overwhelms 
Ca-enhancement and makes Fe5270 weaker on the metal-rich side.  
The top right panel of Figure 9 shows that Fe5270 also decreases 
with blue HB stars on the metal-poor side ([Fe/H] $<$ $-$1.0) 
by 0.2 \AA\ at 14 Gyr.  The bottom left panel of Figure 9 depicts 
that Fe5270 increases in general with a bottom-heavy IMF, particularly 
on the metal-rich side ([Fe/H] $\geq$ $-$1.0).  The bottom right 
panel shows that the effect from blue HBs becomes commensurately 
important with top-heavy IMF on the metal-poor side ([Fe/H] $<$ $-$1.0).  

Although the $\alpha$-element enhancement effects on Fe5335 shown 
in the top left panel of Figure 10 is similar to that on Fe5270 in 
the top left panel of Figure 9, it is noted from the middle right 
panel of Figure 10 that the spectral effects do not make Fe5335 
stronger on the metal-poor side as they do for Fe5270.  According to 
Table 1, it is mostly Mg-enhancement which weakens Fe5335; 
Ca-enhancement does not influence Fe5335 significantly as it does Fe5270.  
The top right panel of Figure 10 displays that the HB effect is 
similar, though in less degree, to Fe5270.  Fe5335 diminishes 
by 0.15 \AA\ with blue HB stars at 14 Gyr.  The bottom panels 
illustrate that the IMF effect is similar to Fe5270.  

The top left panel of Figure 11 depicts that the $\alpha$-element 
enhancement effects on Fe5406 is relatively minor compared to that 
on Fe5270 and Fe5335 on the metal-poor side ([Fe/H] $\leq$ $-$1.0).  
From the middle panels of Figure 11, we find that it is the spectral 
effects that cancel out the isochrone effects on the metal-poor side.  
According to Table 1, all the $\alpha$ elements make Fe5406 weaker, 
though the primary element is Mg.  Similar to Fe5270 and Fe5335, 
the top right panel of Figure 11 illustrates that the blue HB stars 
on the metal-poor side make Fe5406 weaker by $\sim$0.1 \AA\ at 14 Gyr.  
The bottom panels display that the IMF effect is similar to 
Fe5270 and Fe5335.

The top left panel of Figure 12 shows that the $\alpha$-enhancement 
on Fe4383 is similar to that on Fe5406 rather 
than Fe5270 and Fe5335.  It is noted from the middle right panel of 
Figure 12 that the spectral effects due to $\alpha$-element enhancement 
grow systematically stronger with increasing metallicity.  According to 
Table 1, all the $\alpha$ elements except Ti make Fe4383 weaker, 
the leading element is Mg.  Similar to Fe5270, Fe5335, and Fe5406, 
the top right panel of Figure 12 displays that the blue HB stars 
on the metal-poor side make Fe4383 weaker by 0.6 \AA\ at 14 Gyr.  
The bottom panels illustrate that the IMF effect is similar to 
Fe5270, Fe5335, and Fe5406.

\section{Comparisons with Milky Way Globular Clusters} 

Having discussed the theoretical outputs of our SSP model 
$\alpha$-enhanced Lick spectral indices in $\S$ 3, we now provide
empirical checks of our models using the Milky Way GCs.  In this work,
we use recent the large, homogeneous dataset of 41 integrated spectra 
of the Milky Way GCs by S05.  There are 40 MWGCs 
listed in Table 1 of S05 and we find that 
NGC 7078\footnote{We use [Fe/H] = $-$2.45 for NGC 7078 following the 
Table 7 of Kraft \& Ivans (2003) based on MARCS models as we describe 
in section 4.3.} 
is also available from their spectra, hence the total sample has 41 GCs, 
some with multiple observations.  We have averaged the measured Lick 
index values when there are multiple observations.  
Figures 13 $-$ 16 display [Fe/H] vs. Lick 
index plots in order to take advantage of independent measurements 
of [Fe/H] and [$\alpha$/Fe] from individual stars in the Milky Way GCs.  
The [Fe/H] values were taken from Table 1 of S05.  Our simple stellar 
population models with reasonable HB morphologies 
are shown for several Lick indices as a function of [Fe/H].  
The blue lines are solar-scaled models with HB stars (ss-wHB), 
while the pink lines are +0.4 dex $\alpha$-enhanced models (aa-wHB; 
both isochrones and spectra are enhanced).  Ages of our models are noted 
next to the $\alpha$-enhanced models: filled circles are for 8 Gyr, 
open squares are for 11 Gyr, and filled squares are for 14 Gyr, 
respectively.  The SSP models were computed at [Fe/H] = $-$2.5, 
$-$2.0, $-$1.5, $-$1.0, $-$0.5, and 0.0 and are connected by straight 
lines in the figures. 

We describe how our SSP model Lick indices of Mg $b$ and Ca4227, 
Balmer lines, and Fe lines are compare with observations of Milky Way 
GCs in sections 4.1, 4.2, and 4.3, respectively.  Comparisons of 
our model {\em metal line indices} with Milky Way GCs are of great 
importance.  This is because age estimation using the diagnostic 
diagram of a metal line index vs. a Balmer line index could be 
misleading unless the calibrations of model {\em metal line indices} 
are carefully considered.   We remind readers again, however, that the 
purpose of this study is not investigating the detailed match 
between models and the observations for every individual GC, but rather 
to look into the overall match for the entire range of [Fe/H] because of 
the caveats (i.e., no blue stragglers, binaries, multiple populations, 
CNONa considerations, etc) of our models as delineated in section 2.

\subsection{Mg $b$ and Ca4227}

From the left panel of Figure 13, it is seen that the overall match 
between MWGCs and our 0.4 dex $\alpha$-enhanced SSP models is good.  
Although it seems that the metal-poor GCs ([Fe/H] $<$ $-$1.0) indicate 
[$\alpha$/Fe] $\sim$ 0.2 - 0.3 dex, it is again not an aim of this 
study to investigate the every individual Milky Way GC's 
location on our model grids.  There is mounting evidence that 
not every $\alpha$-element is equally enhanced in MWGCs and not every 
MWGC is of the similar amount of $\alpha$-element enhancement 
(e.g., Gratton, Sneden, \& Carretta 2004; Pritzl, Venn, \& Irwin 2005).  
Our models represent the 
case when all the $\alpha$ elements are equally enhanced by 0.4 dex.  
Lee et al. (2009) have noted that Mg $b$ is primarily sensitive to 
Mg alone and other elements do not much affect it so that 
[$\alpha$/Fe] = 0.4 dex should be equivalent to [Mg/Fe] = 0.4 for Mg $b$.  

From the right panel of Figure 13, it is indicated that 
[$\alpha$/Fe] ([Ca/Fe]) $\sim$ +0.2 would reproduce the observations better.  
As Pritzl et al. (2005) showed, it may be true that Ca is less enhanced 
compared to Mg in the stars of MWGCs.  
Pipino et al. (2009) recently suggest that 
the observed under-abundance of Ca with respect to Mg could be 
attributed to the different contributions from supernovae Type Ia 
and supernovae Type II to the nucleosynthesis of these two elements.  
The two most metal-rich Milky Way GCs in Figure 13 
are NGC 6553 ([Fe/H] = $-$0.20) and NGC 6528 ([Fe/H] = $-$0.10).  
It seems that both of them indicate a lower amount 
of $\alpha$-enhancement compared to their metal-poor counterparts.  
For NGC 6553, Cohen et al. (1999) and Alves-Brito et al. (2006) 
report [Ca/Fe] = +0.06 and +0.05, respectively.  For NGC 6528, 
Carretta et al. (2001) and Zoccali et al. (2004) report 
[Mg/Fe] = $-$0.04 and $-$0.06, respectively.

\subsection{Balmer Line Indices}

Balmer lines are widely used as an age indicator because of 
their superb temperature sensitivity in stars, tracing 
the temperature of the main-sequence turnoff better than many other indices.  
However, \citet{lee09} and earlier works 
\citep{wor94b,tmk04,lw05,coe07,sch07} 
found that they are also abundance sensitive to some degree.  The 
horizontal-branch effect on H$\beta$ was also investigated 
by Lee et al. (2000).  

Figures 14 and 15 are similar to Figures 1 and 2 of Lee \& Worthey (2005), 
but here we see the combination of isochrone and spectral effects 
of 0.4 dex $\alpha$-element (O, Mg, Si, S, Ca, Ti) enhancement 
after the correction of the stellar library's intrinsic abundance 
patterns.  From the top right panels of Figures 4 $-$ 8, we have 
learned that blue HB stars significantly affect the Balmer lines 
(H$\beta$, H$\gamma$, H$\delta$) for the old stellar populations 
on the metal-poor side.  It is clear from Figures 14 and 15 that 
the HB effect indeed improves the agreement between our models 
and observations.  It is useful to note here that most Milky Way GCs
have ages between $\sim$11-14 Gyr (e.g., Salaris \& Weiss 2002).  

Figure 14 shows that the overall match is rather good although that 
the observations are a little bit weaker than the models, particularly 
those for $-$1.0 $<$ [Fe/H] $<$ $-$0.6.  We can see 
\citet{poo09} for a more detailed discussion, but a few words about 
outliers might be useful.  The biggest outlier is 
NGC 6544 ([Fe/H] = $-$1.38; red open square) in Figure 14, that shows 
significantly weaker H$\beta$ compared to other MWGCs with similar 
metallicity (although only much slightly weaker at H$\gamma$ and 
H$\delta$ in Figure 15).  According to Harris (1996; 2003 updated version) 
MWGC compilation, NGC 6544 has considerably large reddening 
than other GCs in the S05 sample, 
E(B$-$V) = 0.73.  Moreover, Figure 1 of Hazen (1993) shows that there 
is a rather bright foreground star within 2' from the center of 
NGC 6544.  Several MWGCs with $-$1.0 $<$ [Fe/H] $<$ $-$0.6 
also show some mismatches against models.  
Among the other clusters in the sample, 
it is interesting to find that NGC 6388 ([Fe/H] = $-$0.68; blue open 
circle) and NGC 6441 ([Fe/H] = $-$0.65; green open circle), which have 
unusually sizable population of blue HB stars 
for their metallicities (Rich et al. 1997),
have stronger H$\beta$ features than 47 Tuc (([Fe/H] = $-$0.70; 
red open circle), which has a purely red HB.  

For two most metal-rich MWGCs in the sample, a significant population of 
blue stragglers is seen in NGC 6553 ([Fe/H] = $-$0.20) (Beaulieu et al. 2001; 
Zoccali et al. 2001) and in NGC 6528 ([Fe/H] = $-$0.10) (Brown et al. 2005).  
They may explain rather strong H$\beta$ from NGC 6553 and NGC 6528 compared to 
those H$\beta$-weak MWGCs with $-$1.0 $<$ [Fe/H] $<$ $-$0.6.  

The overall agreement
between the $\alpha$-enhanced SSP models of H$\gamma$ and H$\delta$ 
Lick indices and the MWGCs in Figure 15 are 
much better than what we found for H$\beta$, as shown in Figure 14.
However, it is the case that the $\alpha$-enhanced models differ 
less from the scaled-solar models for H$\gamma$ and H$\delta$ than
they did for H$\beta$.  
NGC 6544 ([Fe/H] = $-$1.38; red open square) 
is still an outlier but not to the great extent that it is for H$\beta$.  
As with H$\beta$, NGC 6388 and NGC 6441, with sizable population of 
blue HB stars, show relatively stronger H$\gamma$ and H$\delta$ 
within a small metallicity range (again, these two and 47 Tuc are marked 
with different colored circles for comparison), particularly at H$\gamma$.  

The top and middle panels of Figures 5 $-$ 8 illuminated that 
the isochrone and spectral effects of $\alpha$-element enhancement 
on H$\gamma$ and H$\delta$ go in the opposite direction and 
their effects are cancelled out on the metal-poor side, exposing mostly 
the HB effect alone.  Among four H$\gamma$ and H$\delta$ 
indices, H$\gamma_A$ has the broadest dynamic range ($\sim$8 \AA\ 
difference between NGC 7078 with [Fe/H] = $-$2.45 and NGC 6528 with 
[Fe/H] = $-$0.10).  The bottom panels of Figure 15 display that 
H$\gamma_F$ and H$\delta_F$ are very similar to H$\gamma_A$ and 
H$\delta_A$ in the upper panels, except for a narrower dynamic range.

\subsection{Iron Line Indices}

In Figure 16, we see that the overall match between 
Lick iron indices and the MWGCs is generally good although 
not superb in Fe5270 particularly around [Fe/H] $\sim$ $-$1.0.  
One of the culprits, is perhaps, the spectral effect of Ca 
as we described in section 3.3.   
Isochrone effects from individual element need to be investigated 
because $\alpha$-enhanced isochrone effects become significant 
near [Fe/H] = $-$1.0.  Dotter et al. (2007a) investigated 
individual elemental effects on the isochrones only at the solar 
metalicity.  The dynamic range 
of Fe5406, which is claimed to be the least sensitive 
to every element except iron among 
Lick iron indices (Lee et al. 2009; Percival et al. 2009), is only
slightly smaller ($\sim$1.6 \AA) compared to that of Fe5270 
($\sim$2.4 \AA) and Fe5335 ($\sim$2.0 \AA).  

Despite the use of high-resolution spectra of individual stars,
the discrepancy of [Fe/H] estimation can be quite large among different
studies even though their internal uncertainties are often fairly small.   
For example, Barbuy et al. (1999) found NGC 6553 to have [Fe/H] = $-$0.55
while Carretta et al. (2001) found [Fe/H]=$-$0.06.  
To demonstrate the effect of changing the [Fe/H] scale we present
Figure 17, which is the same as Figure 16, but the [Fe/H] values of MWGCs 
are from Harris (1996) compilation\footnote{We use his 2003 version, 
http://physwww.mcmaster.ca/~harris/mwgc.dat} instead of Table 1 of 
S05.  Compared to Figure 16, the matches are not 
very favorable especially at $-$1.0 $<$ [Fe/H] $<$ $-$0.5.  

In Figure 18, we compare [Fe/H] values in Table 1 of S05 
with several different [Fe/H] compilations available in the literature.  
They are Harris (1996; 2003 version), Zinn \& West (1984), 
Carretta \& Gratton (1997), and Kraft \& Ivans (2003).  The top left 
panel of Figure 18 compares [Fe/H] values in Table 1 of S05 
with Harris 2003 compilation.  40 MWGCs except NGC 7078 are displayed here.  
The middle left panel shows the comparison 
between S05 and Zinn \& West (ZW84) [Fe/H] in Table 7 
of Kraft \& Ivans (2003).  29 MWGCs are shown here.  
By comparing these two, we notice that the 
Harris 2003 compilation generally follows the ZW84 [Fe/H].  We illustrated 
in Figure 17 that our SSP models do not go very well along with [Fe/H] of 
Harris 2003 compilation.  

The top right panel of Figure 18 compares [Fe/H] values of S05 with 
that of Carretta \& Gratton (CG97)\footnote{They are listed in 
Table 2 of Rutledge, Hesser, \& Stetson (1997).} estimated by 
Rutledge, Hesser, \& Stetson (1997).  30 MWGCs are shown 
here\footnote{NGC 6553 is additionally listed in Table 2 of 
Rutledge et al. (1997) compared to Table 7 
of Kraft \& Ivans (2003).}.  It seems that there are about 
0.18 dex\footnote{This is noted in footnote `d' in Table 1 of 
S05.  In the literature, we note that the very recent [Fe/H] 
estimation of NGC 6388 by Carretta et al. (2007) is $-$0.44 and 
that of NGC 6441 by Gratton et al. (2007) is $-$0.34 
compared to $-$0.68 and $-$0.65, 
respectively from Table 1 of S05.} 
systematic differences between them.  From the comparison between 
middle right panel (Kraft \& Ivans 2003 [Fe/H] estimation based upon 
MARCS models) and bottoms panels (left: Kraft \& Ivans 2003 [Fe/H] 
estimation based upon Kurucz models with convective overshooting, 
right: that without convective overshooting) of Figure 18, it seems that 
the [Fe/H] values in Table 1 of S05 correlate 
very well with [Fe/H] in Table 7 of Kraft \& Ivans (2003) based on 
MARCS models.

\section{Summary and Conclusions}  

We have presented scaled-solar and $\alpha$-enhanced SSP 
models as for several Lick/IDS indices using the \citet{dot07b} stellar 
evolution models combined with high resolution synthetic spectra. 
A detailed systematic variation of HB morphology 
with age and metallicity is incorporated in order to represent 
the Milky Way GCs.  Furthermore, we have
investigated the effect of IMF variation on the Lick indices.  
It is realized that, among the $\alpha$ elements, Mg significantly
influences the spectral effects on Balmer and iron line indices.  
From the comparison of our models of Mg $b$ and Ca4227 with the
Schiavon et al. (2005) dataset (Figure 13), we can confirm an
enhancement of about 0.4 dex for Mg 
and about 0.2 dex for Ca, in agreement with the 
measured amounts from the high resolution spectra of individual 
stars in MWGCs (Pritzl, Venn, \& Irwin 2005).  Moreover, we note 
that our SSP model Lick iron indices comply with 
the Kraft \& Ivans (2003) [Fe/H] based upon MARCS models.  
Balmer lines, particularly H$\gamma$ and H$\delta$, 
of MWGCs are also well reproduced by our $\alpha$-enhanced models 
not only because of the combination of isochrone and spectral 
effects but also because of our reasonable HB treatment.  Finally, 
the investigation of IMF variations on Lick indices reveals that 
a giant-dominant IMF can be ruled out because the Balmer lines 
are too strong on the metal-poor side to be compatible with 
observations.

\acknowledgements 

We are grateful to the anonymous referee for her/his 
constructive report that improves our presentation.  
Support for this work was provided by the NASA through grant HST-GO-11083 
and by the NSF through grant AST-0307487, 
the New Standard Stellar Population Models (NSSPM) project.


\begin{deluxetable}{lrrrrrrrr}
\tablecolumns{9}
\tabletypesize{\small}
\tablecaption{Spectral Effects on Selected Lick Indices at 11 Gyr \label{tab:index1}}
\tablehead{
\colhead{Index} & \colhead{Index} &  \colhead{$\Delta$I} 
& \colhead{$\Delta$I}  & \colhead{$\Delta$I} 
& \colhead{$\Delta$I} & \colhead{$\Delta$I} &
\colhead{$\Delta$I} 
& \colhead{$\Delta$I} \\ 
\colhead{Name} & \colhead{Value} & \colhead{(O)} 
& \colhead{(Mg)} & \colhead{(Si)}
& \colhead{(S)}  & \colhead{(Ca)} & \colhead{(Ti)} 
& \colhead{($\alpha$)}  \\ 
}

\startdata

\sidehead{ [Fe/H] = $-$1.0}

      Mg $b$    &    0.750 & 0.023 &  1.010 & -0.067 &  0.000 & 0.008 &  0.000 &  0.975 \\
      Ca4227   &    0.288 & 0.064 &  0.012 &  -0.004 & -0.001 & 0.352 &  0.005 &  0.431 \\
     H$\beta$   &    2.228 & 0.013 & -0.047 &  0.011 & 0.002 &  -0.004 & 0.052 &  0.024 \\
  H$\gamma_A$   &   -1.713 & 0.246 &  0.408 &  0.213 & 0.001 &  -0.060 & -0.333 & 0.476 \\
  H$\gamma_F$   &    0.393 & 0.148 &  0.193 &  0.114 & 0.000 &  -0.030 & -0.008 & 0.419 \\
  H$\delta_A$   &    1.122 &  0.093 &  0.225 &  0.209 & 0.001 &  0.079 & -0.100 & 0.508 \\
  H$\delta_F$   &    1.364 &  0.003 &  0.054 &  0.262 & 0.001 &  0.089 & -0.073 & 0.333 \\
      Fe5270   &    1.621 & -0.018 & -0.063 & -0.016 &  0.001 &  0.138 &  0.050 & 0.095 \\
      Fe5335   &    1.432 & -0.009 & -0.057 & -0.013 &  0.000 &  0.007 &  0.024 & -0.049 \\
      Fe5406   &    0.888 & -0.006 & -0.029 & -0.013 &  0.000 & -0.005 &  -0.030 &  -0.081 \\
      Fe4383   &    2.611 & -0.068 & -0.235 & -0.176 &  -0.001 & -0.112 &  0.121 &  -0.472 \\

\sidehead{ [Fe/H] = 0.0}

      Mg $b$    &    4.021 & 0.089 &  1.775 & -0.162 &  0.000 & -0.018 &  0.005 &  1.689 \\
      Ca4227   &    1.357 & 0.213 &  0.007 &  -0.032 &  0.001 & 0.841 &  0.000 &  1.028 \\
     H$\beta$   &    1.809 & 0.040 & -0.220 & -0.021 & 0.000 &  0.005 & 0.064 &  -0.134 \\
  H$\gamma_A$   &   -6.245 & 0.614 &  1.165 &  0.447 & 0.001 &  -0.099 & -0.681 & 1.445 \\
  H$\gamma_F$   &   -1.679 & 0.338 &  0.452 &  0.189 & 0.001 &  -0.064 & -0.045 & 0.870 \\
  H$\delta_A$   &   -2.603 &  0.290 &  0.861 &  0.531 & 0.001 &  0.397 & -0.217 & 1.862 \\
  H$\delta_F$   &   -0.056 & -0.019 &  0.184 &  0.552 & 0.000 &  0.219 & -0.150 & 0.787 \\
      Fe5270   &    3.183 & -0.100 & -0.240 & -0.068 &  0.004 &  0.173 &  0.044 & -0.187 \\
      Fe5335   &    2.958 & -0.042 & -0.236 & -0.075 &  0.000 & -0.030 &  0.025 & -0.356 \\
      Fe5406   &    1.847 & -0.032 & -0.126 & -0.052 &  0.000 & -0.030 &  -0.028 &  -0.268 \\
      Fe4383   &    5.838 & -0.076 & -0.729 & -0.434 &  0.000 & -0.328 &  0.094 &  -1.472 \\

\enddata
\tablecomments{1. The units are \AA\ of equivalent width.  2. The second column
               is the index of solar-scaled models without HB stars.  
               3. $\Delta$ I = index of 
               each element-enhanced by 0.4 dex at fixed [Fe/H]   
               $-$ index of solar-scaled model, both without HB stars.  
               4. $\alpha$ in the last column is the case that all of 
               O, Mg, Si, S, Ca, and Ti are enhanced by 0.4 dex at fixed [Fe/H] 
               without HB stars.}
\end{deluxetable}


\clearpage

\begin{figure}
\epsscale{1.}
\plotone{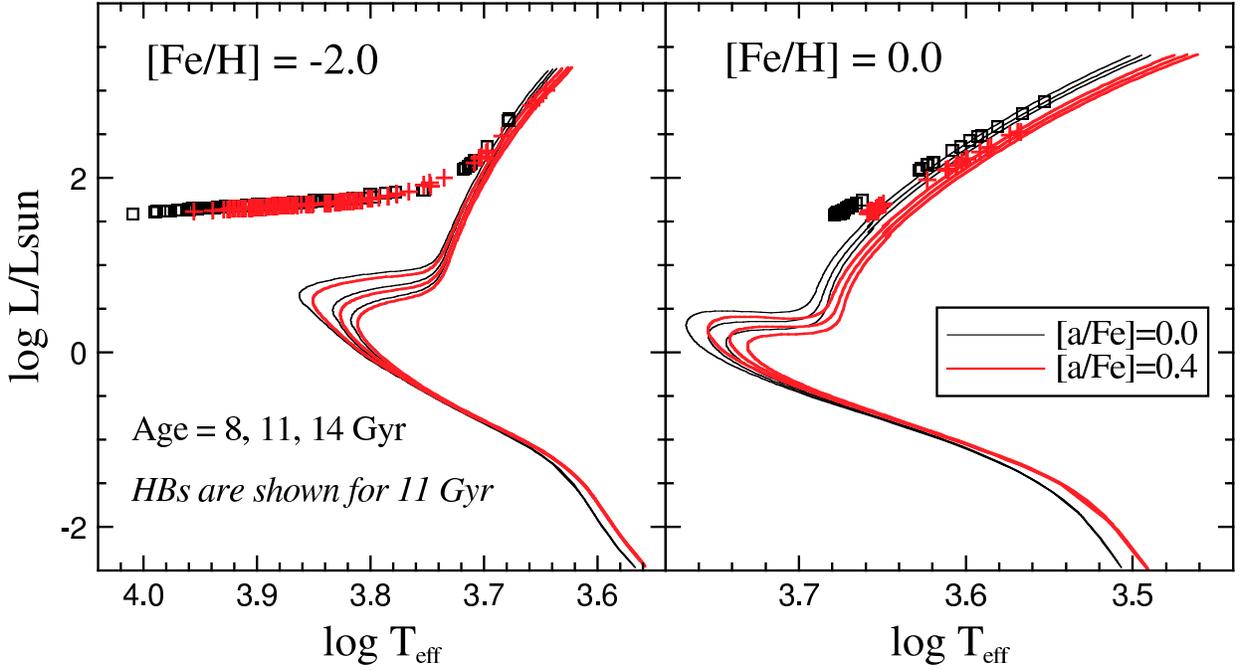}
\caption{H-R diagram comparisons are illustrated between solar-scaled 
and 0.4 dex $\alpha$-element enhanced isochrones at 8, 11, and 14 Gyr.  
The left panel shows them at [Fe/H] = $-$2.0, while the right panel 
displays them at [Fe/H] = 0.0.  The horizontal-branch (HB) morphologies 
are only shown for 11 Gyr.  The small squares are post-ZAHB (HB + AGB) 
stars for the solar-scaled, while the crosses are 
that for the $\alpha$-enhanced.  The $\alpha$-enhanced isochrones and 
post-ZAHB stars here at fixed [Fe/H] are generally cooler than that of the 
solar-scaled.  Moreover, it is noted that the $\alpha$-element enhancement 
effects are more significant on the metal-rich side.
\label{fig01}}
\end{figure}

\begin{figure}
\epsscale{.5}
\plotone{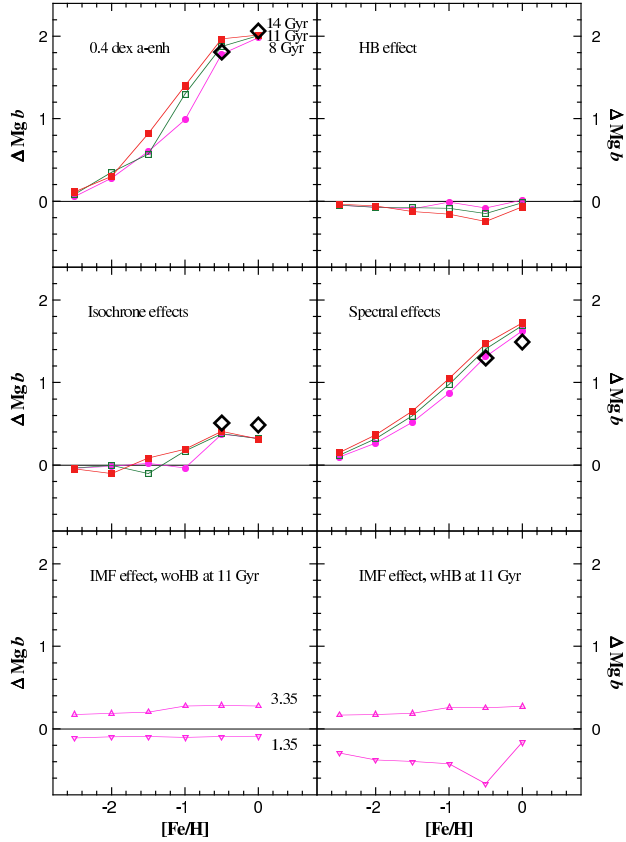}
\caption{The residuals of our simple stellar population (SSP) models 
for Mg $b$ are illuminated.  
In the top left panel, the differences between 0.4 dex $\alpha$-enhanced 
models (both isochrones and spectra are enhanced) without HB stars and 
that of the solar-scaled (aa-woHB $-$ ss-woHB) are displayed.  Ages of 
our SSP models are denoted as filled circles for 8 Gyr, open 
squares for 11 Gyr, and filled squares for 14 Gyr, respectively.  
Metallicities of our SSP models are given at [Fe/H] = $-$2.5, $-$2.0, 
$-$1.5, $-$1.0, $-$0.5, and 0.0.  The top right panel shows the effect 
of HB stars at the $\alpha$-enhanced models (aa-wHB $-$ aa-woHB).  
The middle panels are same as the top left panel, but the middle left 
panel displays isochrone effects alone without HB 
stars (as-woHB $-$ ss-woHB), while the middle right panel depicts 
spectral effects alone without HB stars (sa-woHB $-$ ss-woHB).  The 
combination of the middle panels corresponds to the top left panel.  
It is demonstrated in the middle panels that both the 
$\alpha$-enhanced isochrones and spectra make Mg $b$ stronger 
as a function of [Fe/H], but the latter are dominant.  The top 
panels show that HB stars do not influence Mg $b$ much compared to 
$\alpha$-element enhancement.  The diamonds are 
Coelho et al. (2007) (see text).  The bottom panels depict the variation 
of IMF effect, 3.35 for dwarf-dominant, 1.35 for giant-dominant, without 
(left) and with HB stars (right) at 11 Gyr for our 0.4 dex 
$\alpha$-enhanced SSP models (see section 2.3 for details).  It is seen 
from the bottom panels that the dwarf-dominant IMF generally makes 
Mg $b$ stronger. 
\label{fig02}}
\end{figure}

\begin{figure}
\epsscale{.7}
\plotone{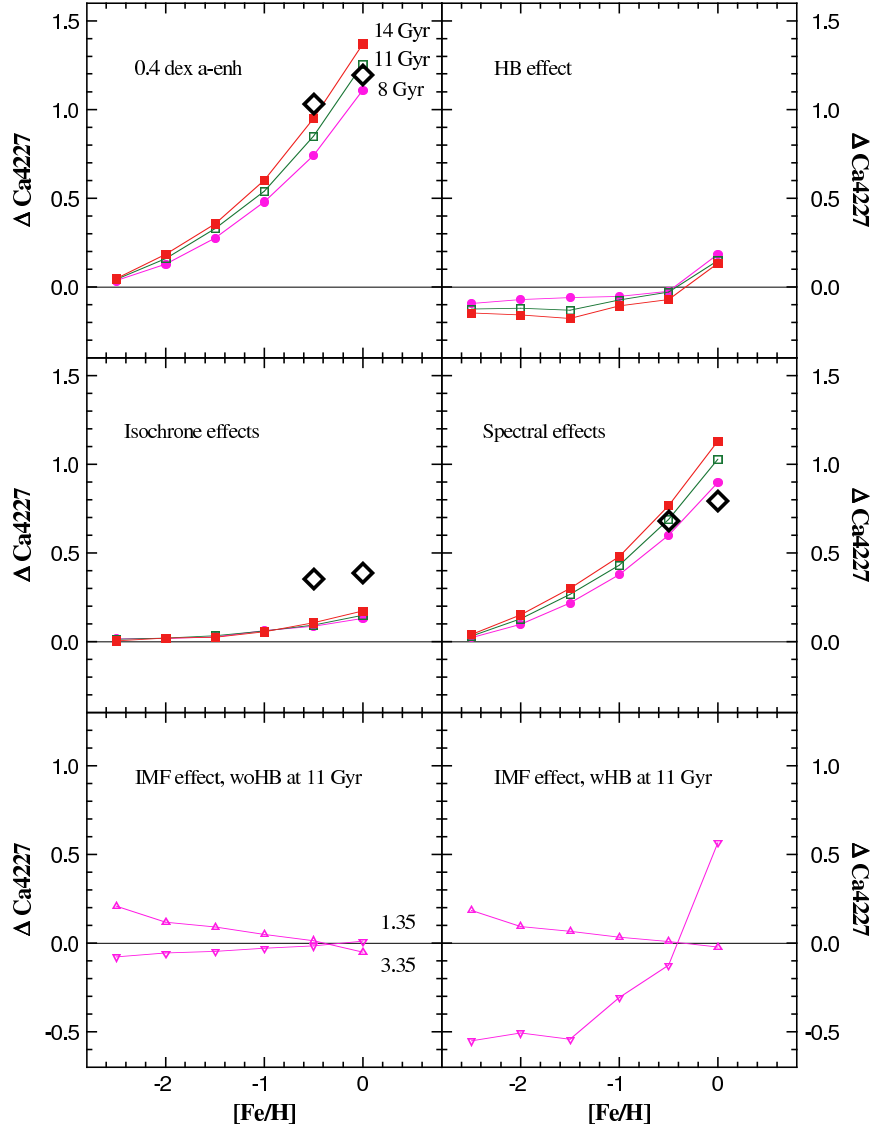}
\caption{Same as Figure 2, but for Ca4227.  The top left panel 
illustrates the combination of the middle panels.  Similar to Mg $b$, 
it is shown in the middle panels that both the $\alpha$-enhanced 
isochrones and spectra make Ca4227 stronger as a function of [Fe/H], 
but the latter are dominant.  Similar to Mg $b$, the top panels 
display that HB stars do not affect Ca4227 much compared to 
$\alpha$-element.  The diamonds are 
Coelho et al. (2007) (see text).  
The bottom panels show that the dwarf-dominant 
IMF generally makes Ca4227 stronger except at the solar metallicity.  
\label{fig03}}
\end{figure}

\begin{figure}
\epsscale{.7}
\plotone{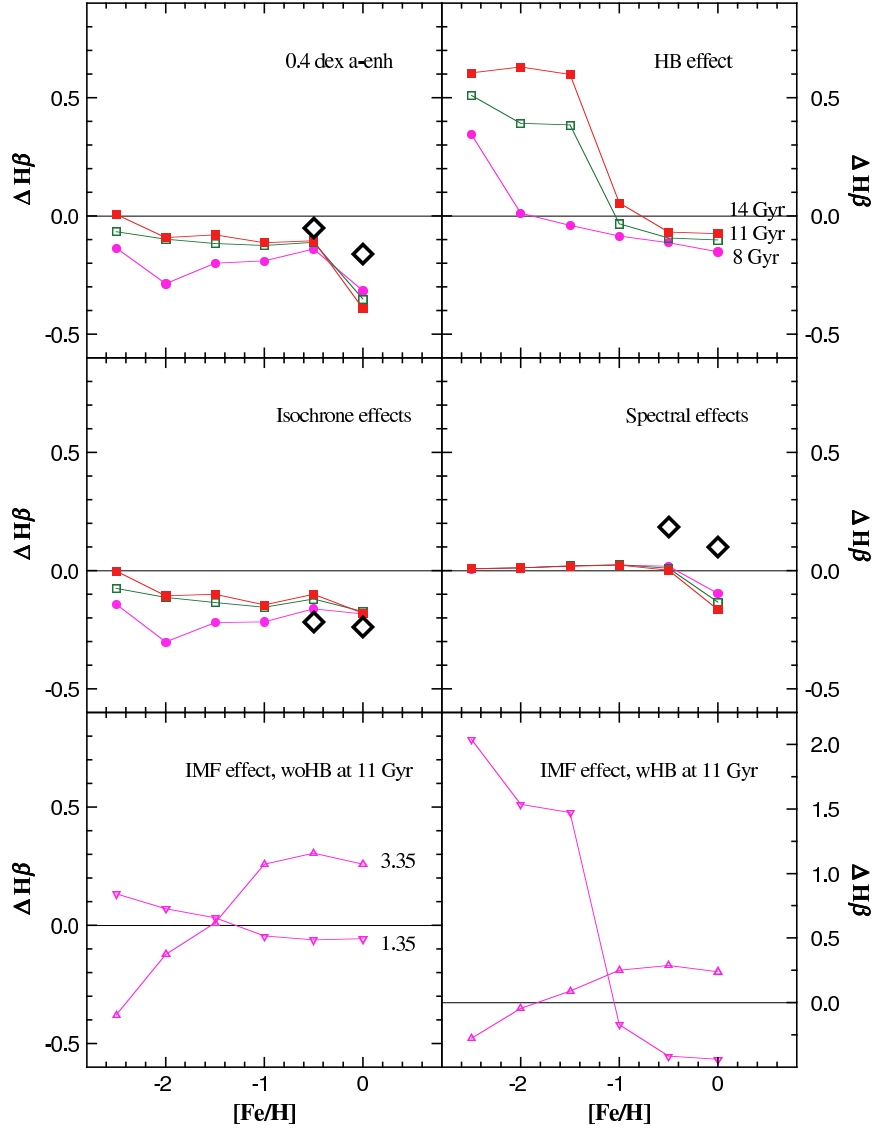}
\caption{Same as Figure 2, but for H$\beta$.  The top left panel 
illustrates the combination of the middle panels.  It is seen from 
the top right panel that H$\beta$ increases as much as 
0.7 \AA\ on the metal-poor side for old stellar populations due to 
blue HB stars.  Also, it is noted from the middle panels that 
the $\alpha$-element enhancement effects are mostly due to the cooler 
isochrones, which decrease the temperature sensitive H$\beta$, 
while the spectral effects are almost negligible except at the 
solar metallicity.  The diamonds are 
Coelho et al. (2007) (see text).  
The bottom panels show that the giant-dominant 
IMF generally makes H$\beta$ stronger on the metal-poor side and 
the vice versa, which is attributed to the temperature sensitivity 
of H$\beta$.  
\label{fig04}}
\end{figure}

\begin{figure}
\epsscale{.7}
\plotone{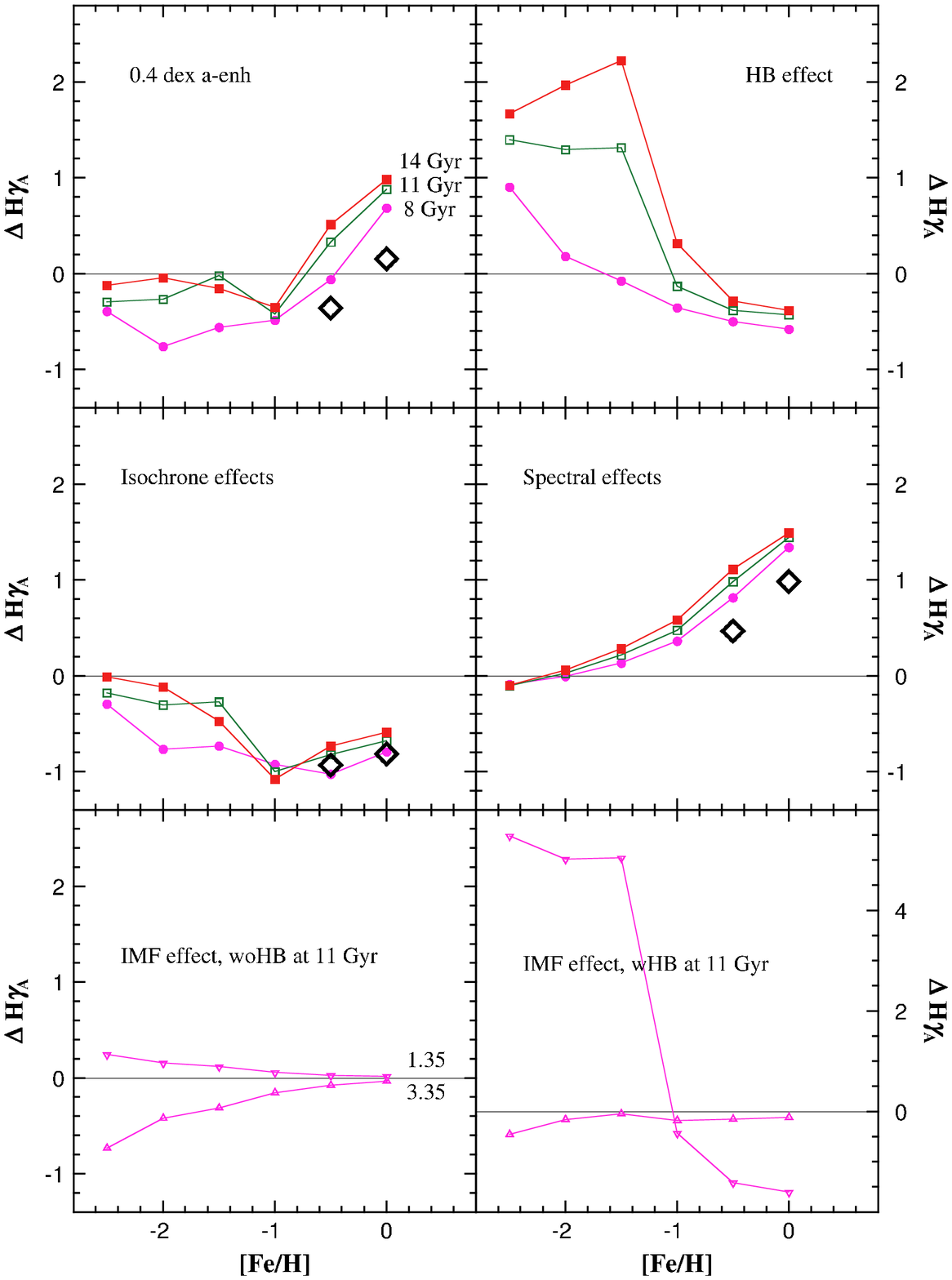}
\caption{Same as Figure 2, but for H$\gamma_A$.  The top left panel 
illustrates the combination of the middle panels.  Similar to H$\beta$, 
the top right panel shows that H$\gamma_A$ increases as much as 
2.4 \AA\ on the metal-poor side for old stellar populations due to blue 
HB stars.  Also, it is noted from the middle panels that the isochrone 
and spectral effects due to $\alpha$-element are opposite.  The diamonds 
are Coelho et al. (2007) (see text).  The bottom right panel shows that 
the giant-dominant IMF makes H$\gamma_A$ considerably stronger 
on the metal-poor side mostly because of the blue HB stars.  
\label{fig05}}
\end{figure}

\begin{figure}
\epsscale{.7}
\plotone{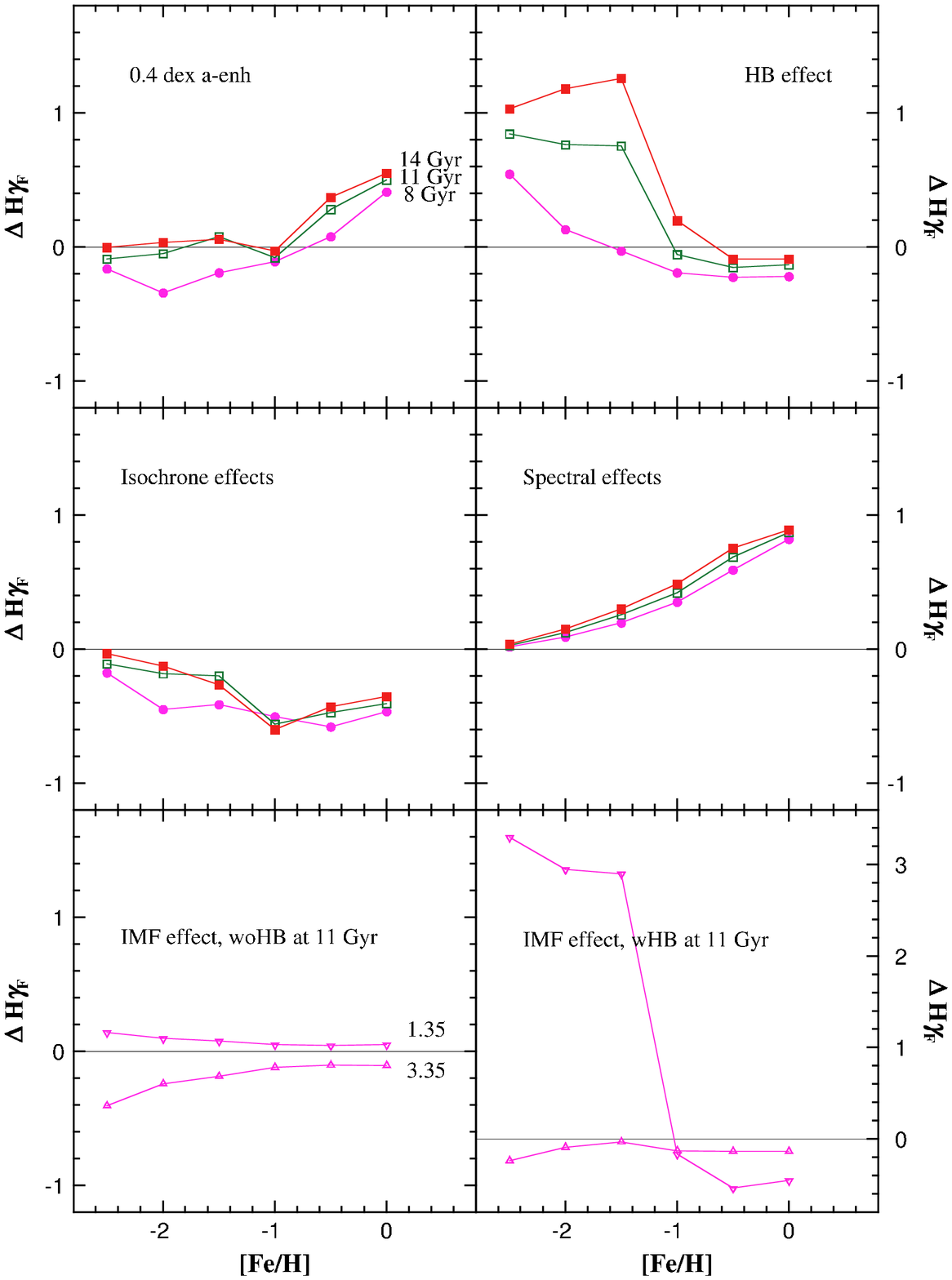}
\caption{Same as Figure 2, but for H$\gamma_F$.  The top left panel 
illustrates the combination of the middle panels.  Similar to H$\gamma_A$, 
the top right panel shows that H$\gamma_F$ increases as much as 1.4 \AA\ 
on the metal-poor side for old stellar populations due to blue HB stars.  
Also, it is noted from the middle panels that the isochrone and spectral 
effects due to $\alpha$-element are opposite.  The bottom right panel 
shows that the giant-dominant IMF makes H$\gamma_F$ considerably
stronger on the metal-poor side mostly because of the blue HB stars. 
\label{fig06}}
\end{figure}

\begin{figure}
\epsscale{.7}
\plotone{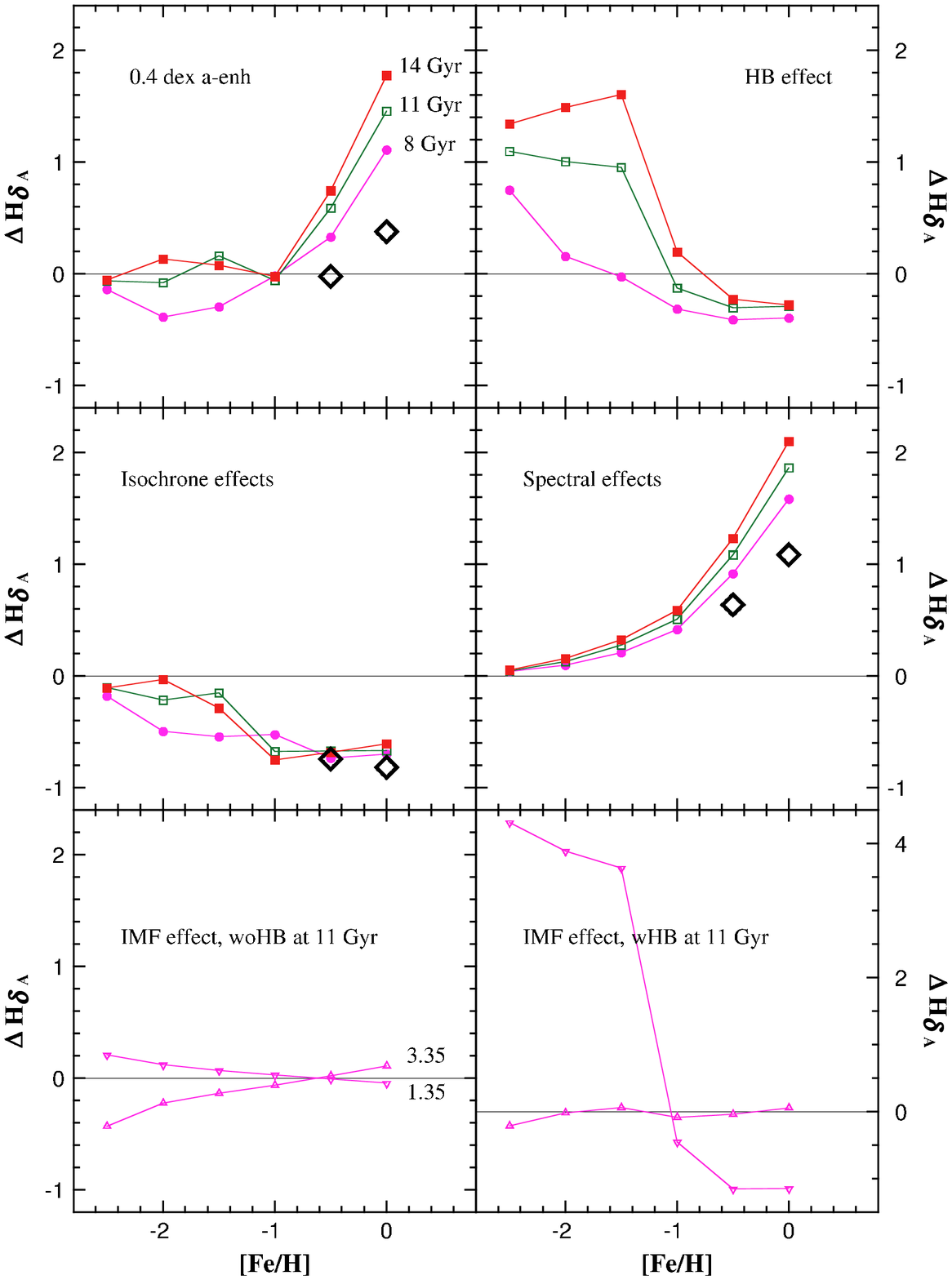}
\caption{Same as Figure 2, but for H$\delta_A$.  The top left panel 
illustrates the combination of the middle panels.  Similar to H$\gamma_A$, 
the top right panel shows that H$\delta_A$ increases as much as 
1.8 \AA\ on the metal-poor side for old stellar populations due to 
blue HB stars.  Also, it is noted from the middle panels that the isochrone 
and spectral effects due to $\alpha$-element are opposite.  
The diamonds are Coelho et al. (2007) (see text).  
The bottom right panel shows that the giant-dominant 
IMF makes H$\delta_A$ considerably stronger on the metal-poor side mostly 
because of the blue HB stars.  
\label{fig07}}
\end{figure}

\begin{figure}
\epsscale{.7}
\plotone{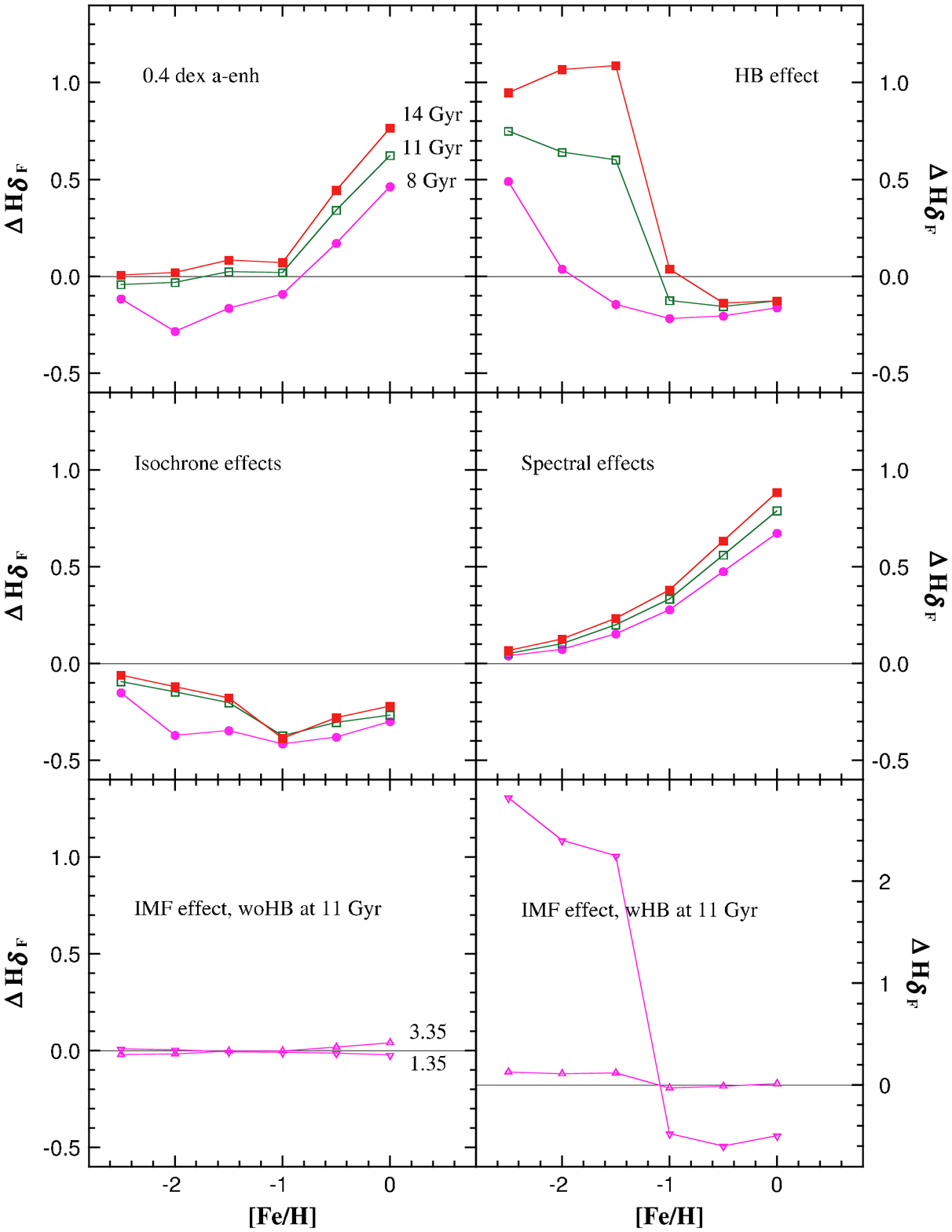}
\caption{Same as Figure 2, but for H$\delta_F$.  The top left panel 
illustrates the combination of the middle panels.  Similar to H$\delta_A$, 
the top right panel shows that H$\delta_F$ increases as much as 1.2 \AA\ 
on the metal-poor side for old stellar populations due to blue HB stars.  
Also, it is noted from the middle panels that the isochrone and spectral 
effects due to $\alpha$-element are opposite.  The bottom right panel 
shows that the giant-dominant IMF makes H$\delta_F$ considerably stronger 
on the metal-poor side mostly because of the blue HB stars.  The bottom 
left panel shows that the IMF effect on H$\delta_F$ is minimal among 
Balmer lines unless we consider HB stars.  
\label{fig08}}
\end{figure}

\begin{figure}
\epsscale{.7}
\plotone{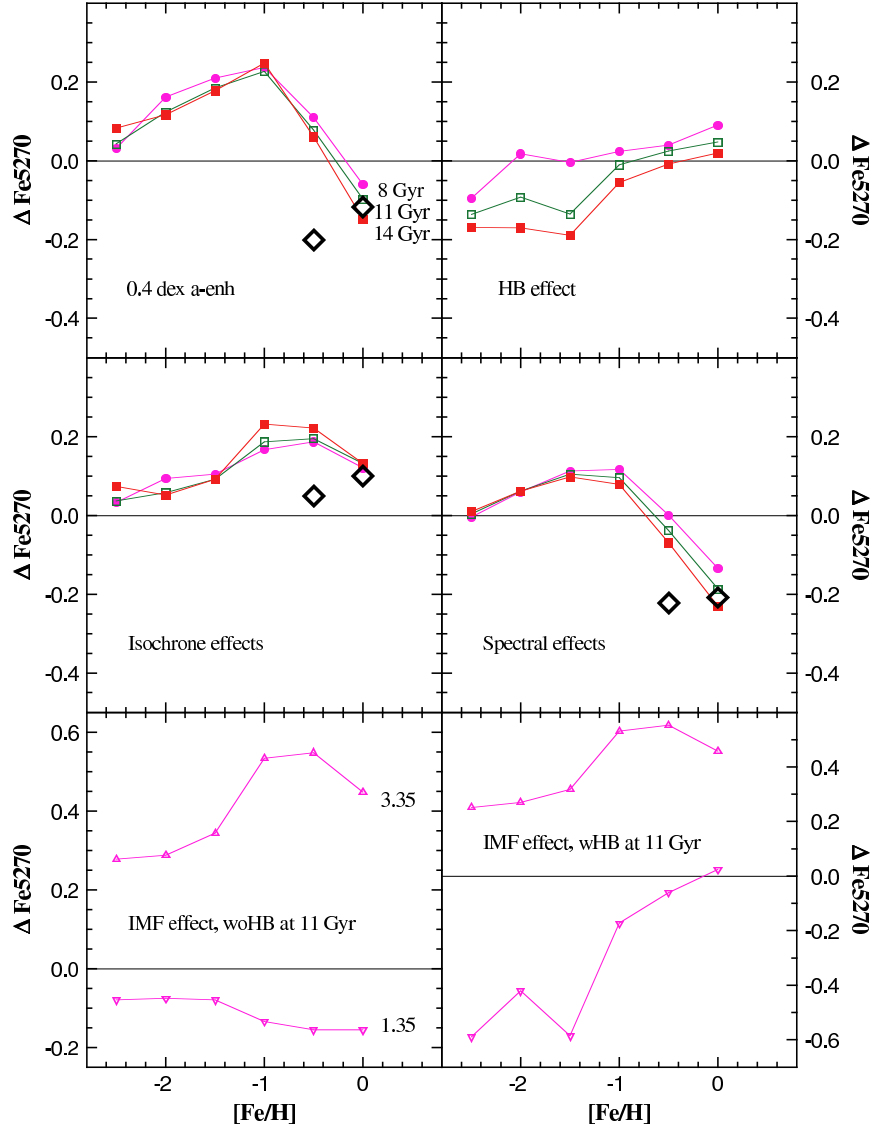}
\caption{Same as Figure 2, but for Fe5270.  The top left panel 
illustrates the combination of the middle panels.  The top right panel 
shows that Fe5270 decreases with blue HB stars on the metal-poor side.  
The middle panels display that both isochrone and spectral effects 
due to $\alpha$-element make Fe5270 stronger on the metal-poor side, 
but the spectral effects near solar metallicity make it weaker.  
The diamonds are Coelho et al. (2007) (see text).  
The bottom panels show that the dwarf-dominant 
IMF makes Fe5270 stronger.  
\label{fig09}}
\end{figure}

\begin{figure}
\epsscale{.7}
\plotone{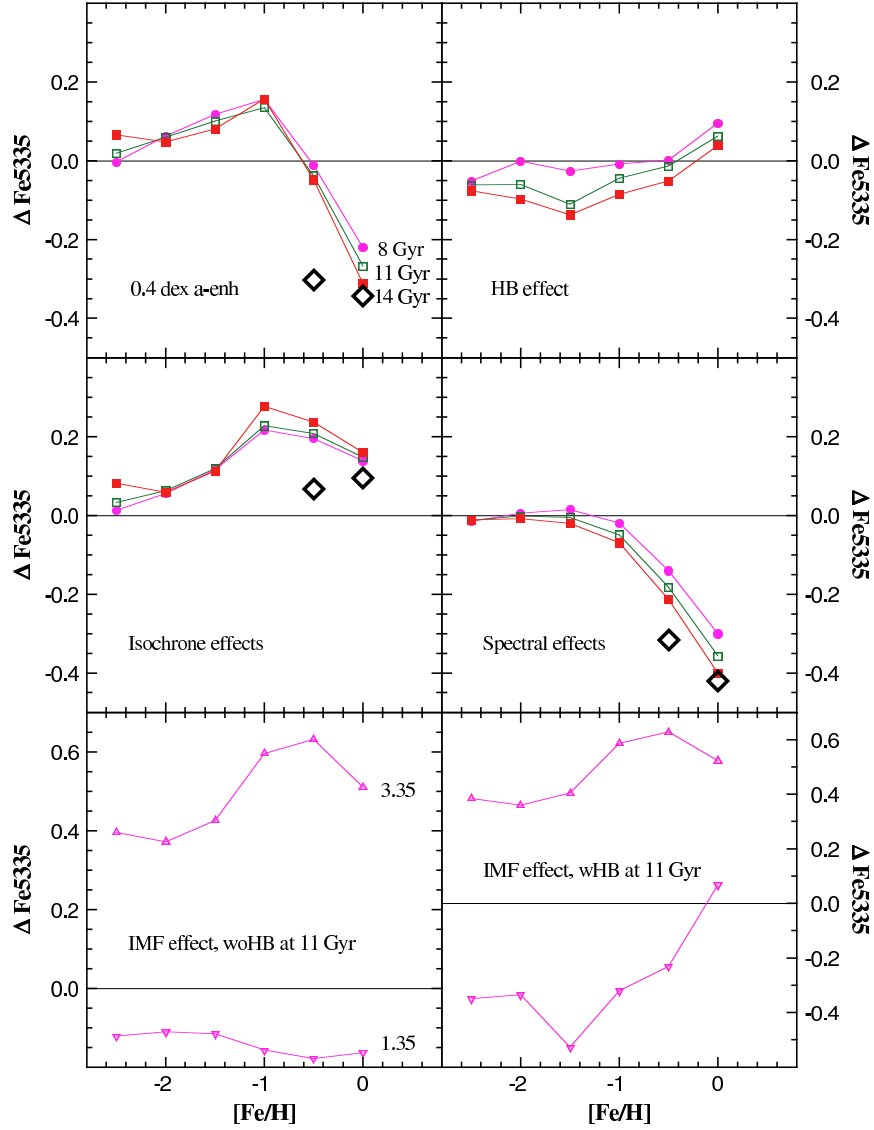}
\caption{Same as Figure 2, but for Fe5335.  The top left panel 
illustrates the combination of the middle panels.  Similar to Fe5270, 
but in less degree, the top right panel shows that Fe5335 decreases 
with blue HB stars on the metal-poor side.  It is noted from 
the middle panels that the isochrone and spectral effects due to 
$\alpha$-element are opposite.  
The diamonds are Coelho et al. (2007) (see text).  The bottom 
panels show that the dwarf-dominant IMF makes Fe5335 stronger.
\label{fig10}}
\end{figure}

\begin{figure}
\epsscale{.7}
\plotone{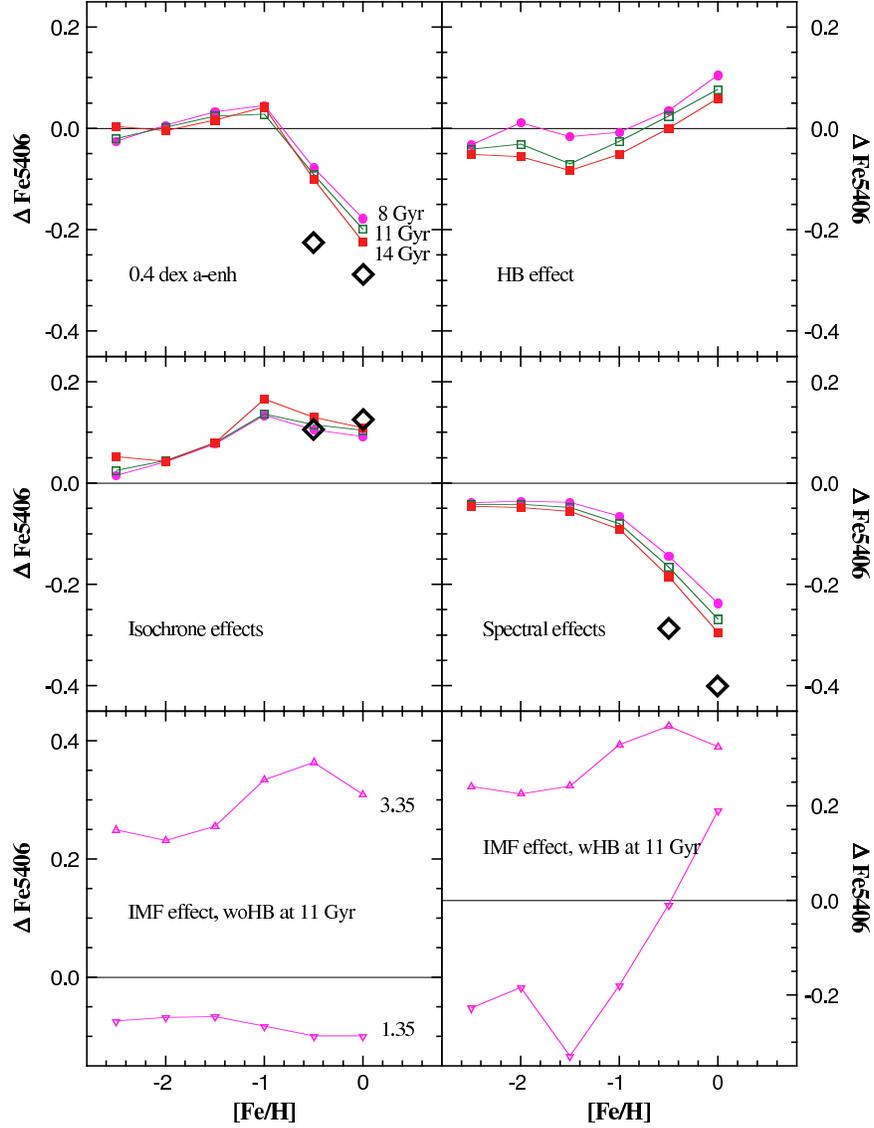}
\caption{Same as Figure 2, but for Fe5406.  The top left panel 
illustrates the combination of the middle panels.  Similar to Fe5270 
and Fe5335, the top right panel shows that HB effect on Fe5406 
decreases with blue HB stars on the metal-poor side.  Compared to Fe5270 
and Fe5335, it is noted from the middle right panel that the 
spectral effects due to $\alpha$-element make Fe5406 weaker even at 
the metal-poor side, [Fe/H] $<$ $-$1.0.  
The diamonds are Coelho et al. (2007) (see text).  
The bottom panels show that the dwarf-dominant IMF makes Fe5406 
stronger.
\label{fig11}}
\end{figure}

\begin{figure}
\epsscale{.7}
\plotone{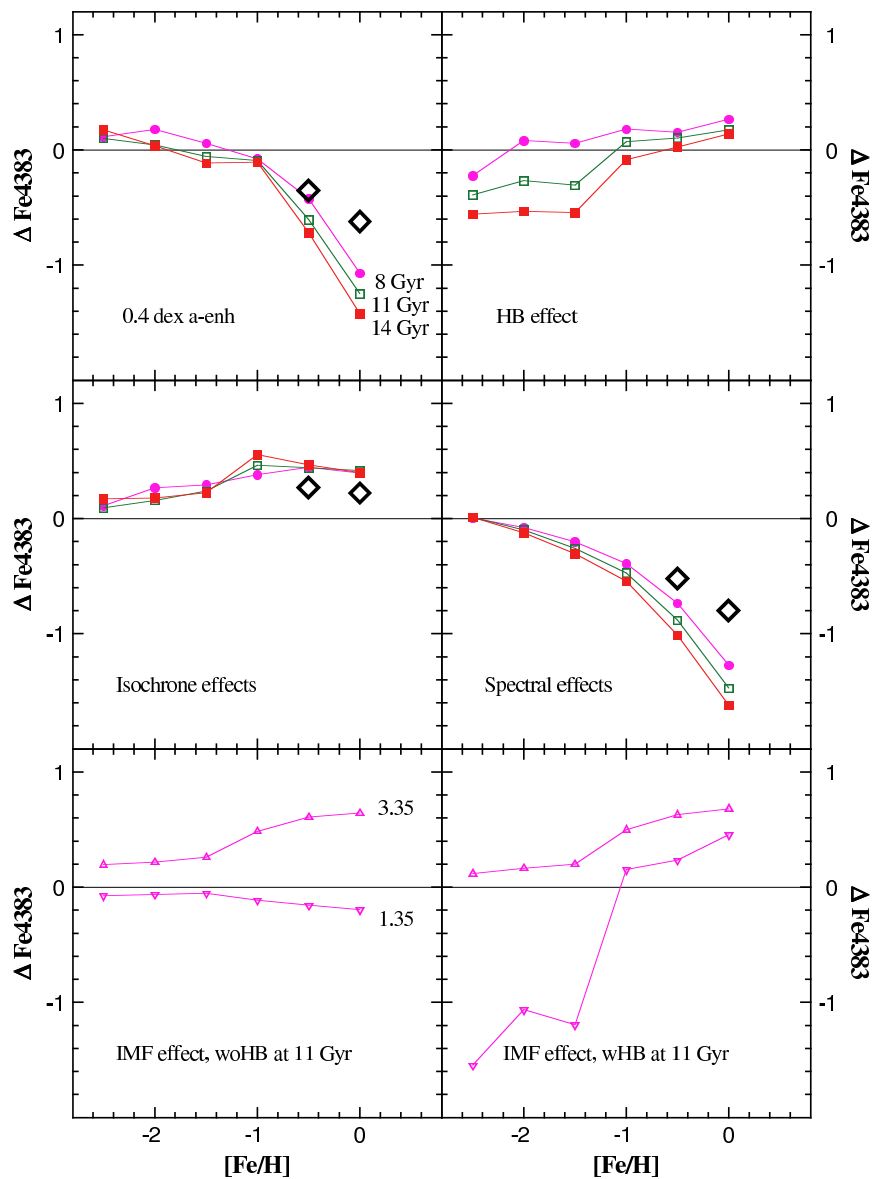}
\caption{Same as Figure 2, but for Fe4383.  The top left panel 
illustrates the combination of the middle panels.  Similar to Fe5270, 
the top right panel shows that Fe4383 decreases with blue HB 
stars on the metal-poor side.  It is noted from the middle right 
panel that the spectral effects due to $\alpha$-element are 
systematically stronger with increasing metallicity.  
The diamonds are Coelho et al. (2007) (see text).  
The bottom panels show that the dwarf-dominant IMF makes 
Fe4383 stronger.
\label{fig12}}
\end{figure}

\begin{figure}
\epsscale{1.}
\plotone{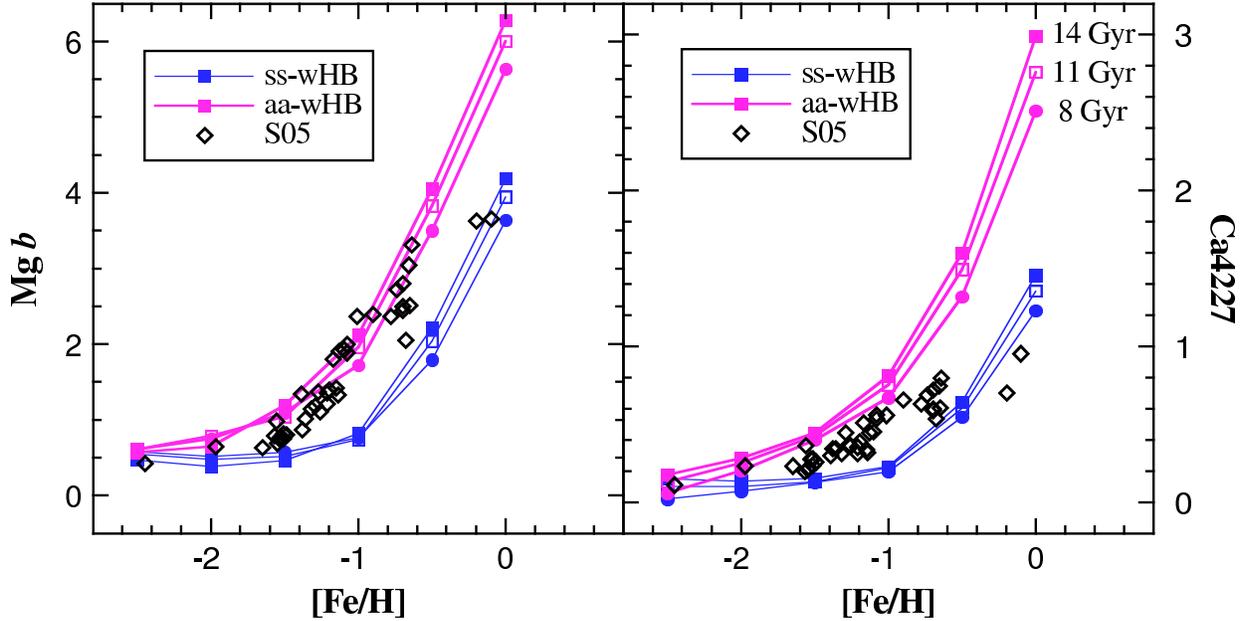}
\caption{Our simple stellar population models (SSP) with realistic 
horizontal-branch (HB) morphologies are shown for Mg $b$ (left) and 
Ca4227 (right) as a function of [Fe/H].  The blue lines are our 
solar-scaled SSP models with HB stars (ss-wHB), while the pink lines 
are that of 0.4 dex $\alpha$-enhanced (aa-wHB; both isochrones and 
spctra are enhanced).  Ages of our models are denoted next to the 
$\alpha$-enhanced models in right panel.  Metallicities of our 
models are given at [Fe/H] = $-$2.5, $-$2.0, $-$1.5, $-$1.0, $-$0.5, 
and 0.0.  The diamonds are 41 Milky Way globular clusters (MWGCs) 
from Schiavon et al. (2005; S05).  The [Fe/H] of MWGCs are also from 
Table 1 of S05.  Because Mg $b$ and Ca4227 are predominantly sensitive 
to Mg and Ca, respectively, among $\alpha$ elements, it is denoted 
here that the MWGCs are, in general, of [Mg/Fe] $\sim$ 0.4 dex and 
[Ca/Fe] $\sim$ 0.2 dex.  The two most metal-rich MWGCs in the sample, 
NGC 6553 ([Fe/H]=$-$0.20) and NGC 6528 ([Fe/H]=$-$0.10), however, 
seem to indicate the less amount of enhancement of $\alpha$-element 
compared to the metal-poor counterparts (see text for details).  
\label{fig13}}
\end{figure}

\begin{figure}
\epsscale{1.}
\plotone{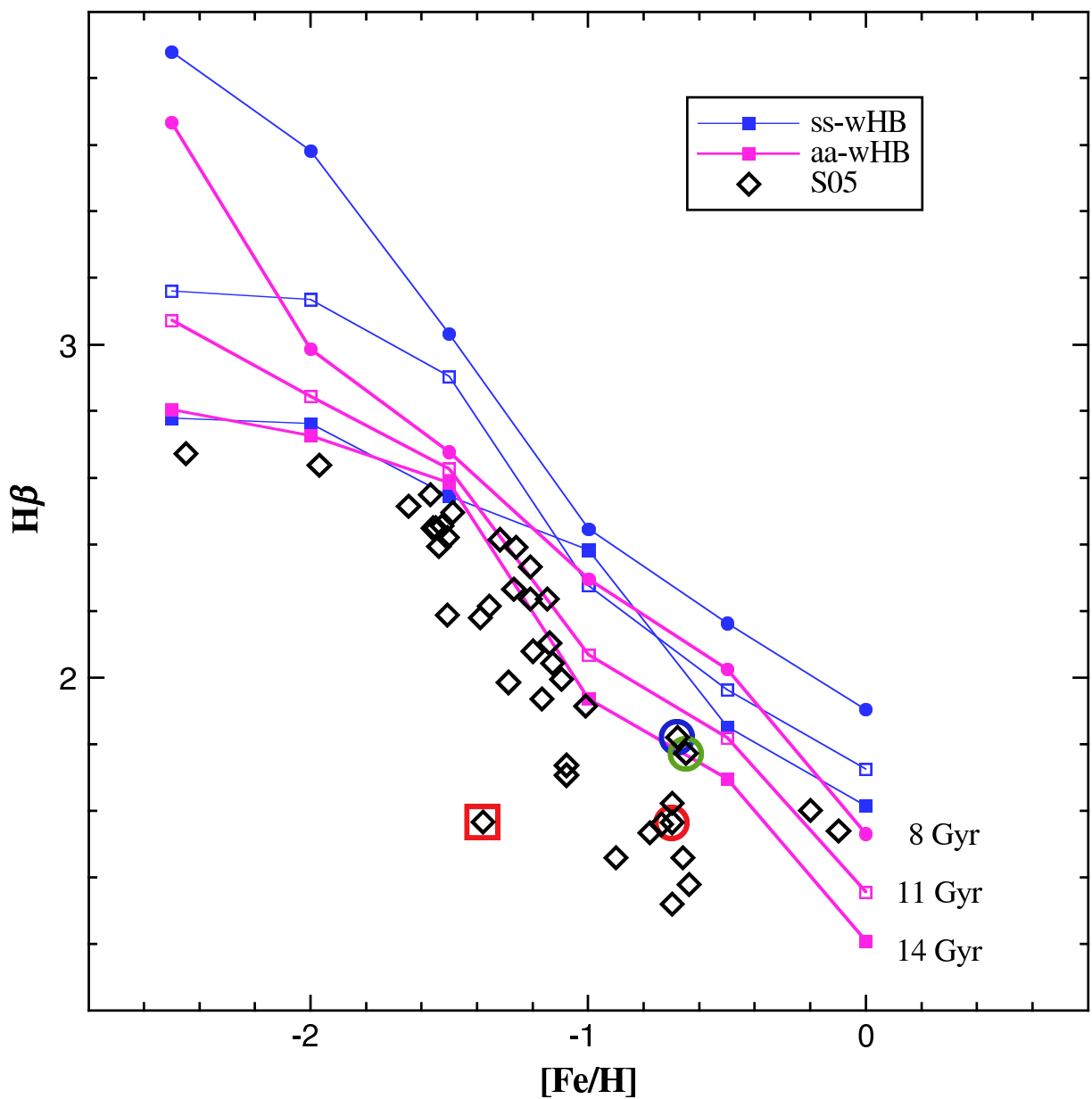}
\caption{H$\beta$ as a function of [Fe/H], with lines and symbols 
as in Figure 13, except for some clusters marked with big symbols.  
The latter are NGC 6544 ([Fe/H] = $-$1.38; red open square), which is 
considerably weaker in H$\beta$ compared to other MWGCs with similar 
metallicity, NGC 6388 ([Fe/H] = $-$0.68; blue open circle) and 
NGC 6441 ([Fe/H] = $-$0.65; green open circle) are of the strongest 
H$\beta$ with sizable numbers of blue HB 
stars compared to 47 Tuc ([Fe/H] = $-$0.70; red open circle) that has a 
pure red clump (see text for details).  
\label{fig14}}
\end{figure}

\begin{figure}
\epsscale{1.}
\plotone{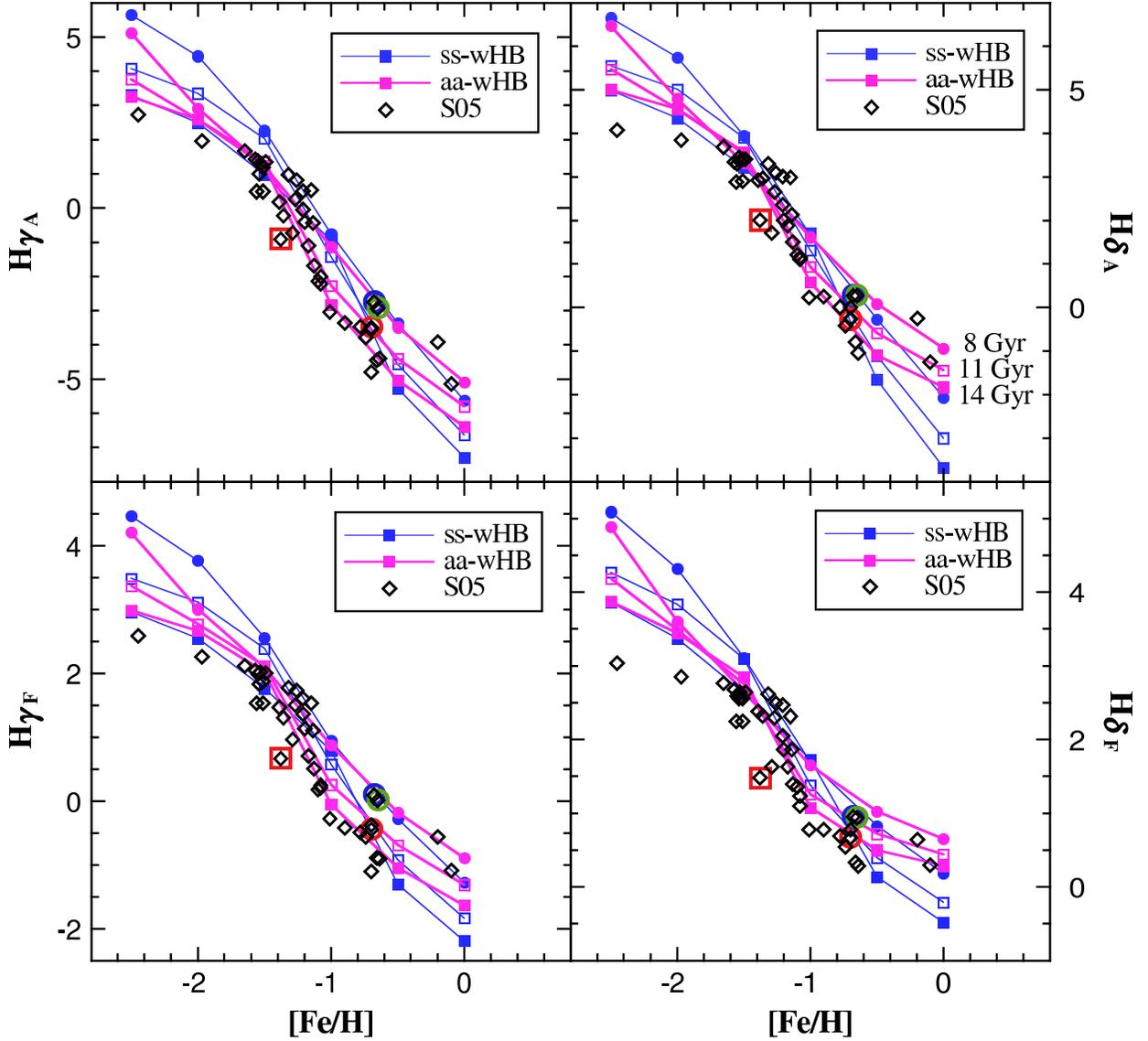}
\caption{Similar to Figure 14, but for H$\gamma_A$ (upper left), 
H$\delta_A$ (upper right), H$\gamma_F$ (lower left), and H$\delta_F$ 
(lower right).  It seems that the matches between our $\alpha$-enhanced 
SSP models and the MWGCs in H$\gamma$ and H$\delta$ are, in general, 
much better than that in H$\beta$ as shown in Figure 14.  
Unlike H$\beta$, NGC 6544 ([Fe/H] = $-$1.38; red open 
square) does not significantly stand out.  Similar to H$\beta$, NGC 6388 
([Fe/H] = $-$0.68; blue open circle) and NGC 6441 ([Fe/H] = $-$0.65; 
green open circle) are of the stronger H$\gamma$ and H$\delta$ with 
sizable blue HB stars compared to 47 Tuc ([Fe/H] = $-$0.70; red open 
circle) which a strong red clump with basically zero bluer HB stars 
(see text for details).
\label{fig15}}
\end{figure}

\begin{figure}
\epsscale{1.}
\plotone{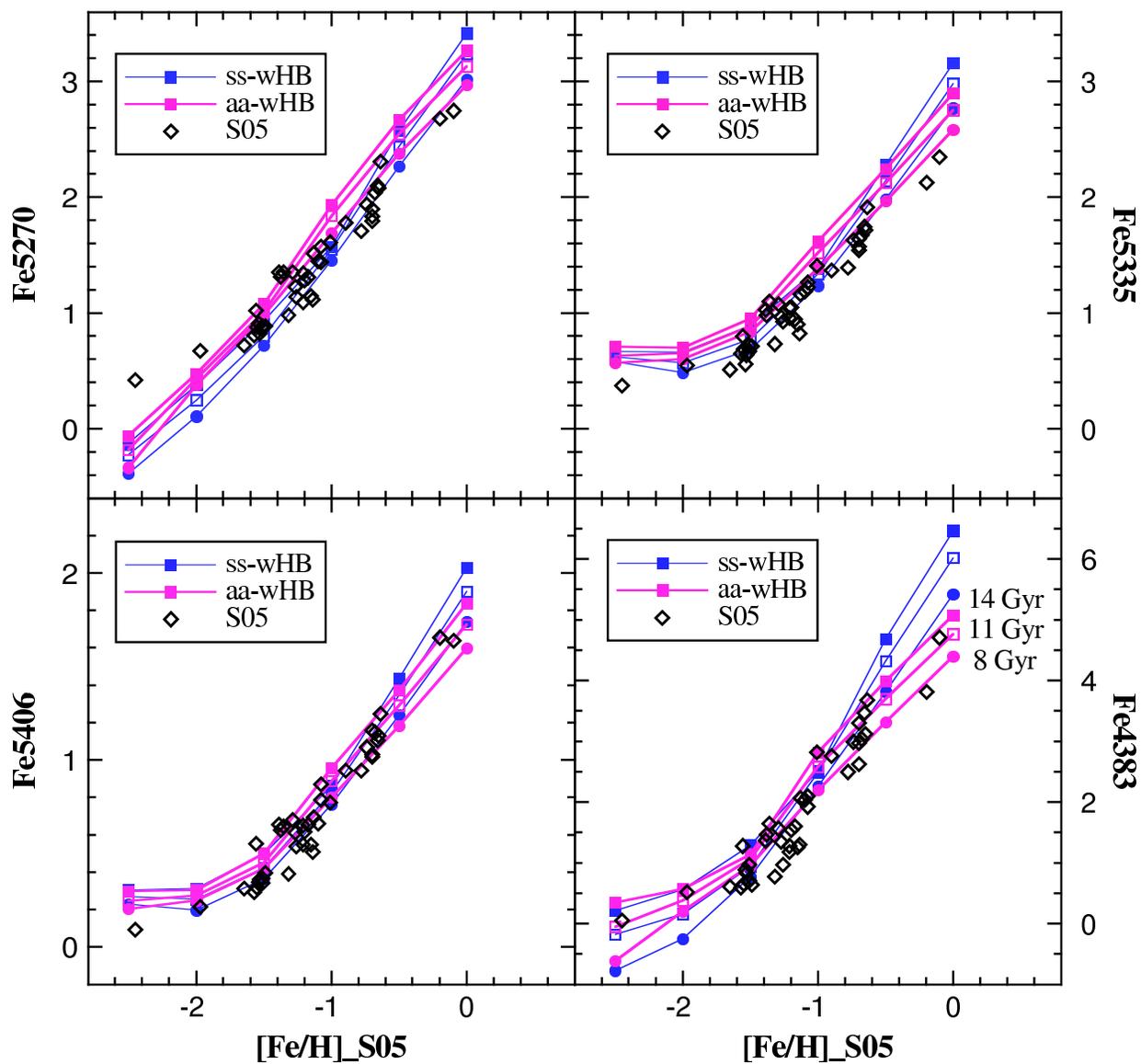}
\caption{Similar to Figure 13, but for Fe5270 (upper left), Fe5335 (upper 
right), Fe5406 (lower left), and Fe4383 (lower right).  It appears that 
the matches between our SSP models and the MWGCs in Lick iron indices are 
generally good, particularly in Fe5406 and Fe4383 (see text for details).  
\label{fig16}}
\end{figure}

\begin{figure}
\epsscale{1.}
\plotone{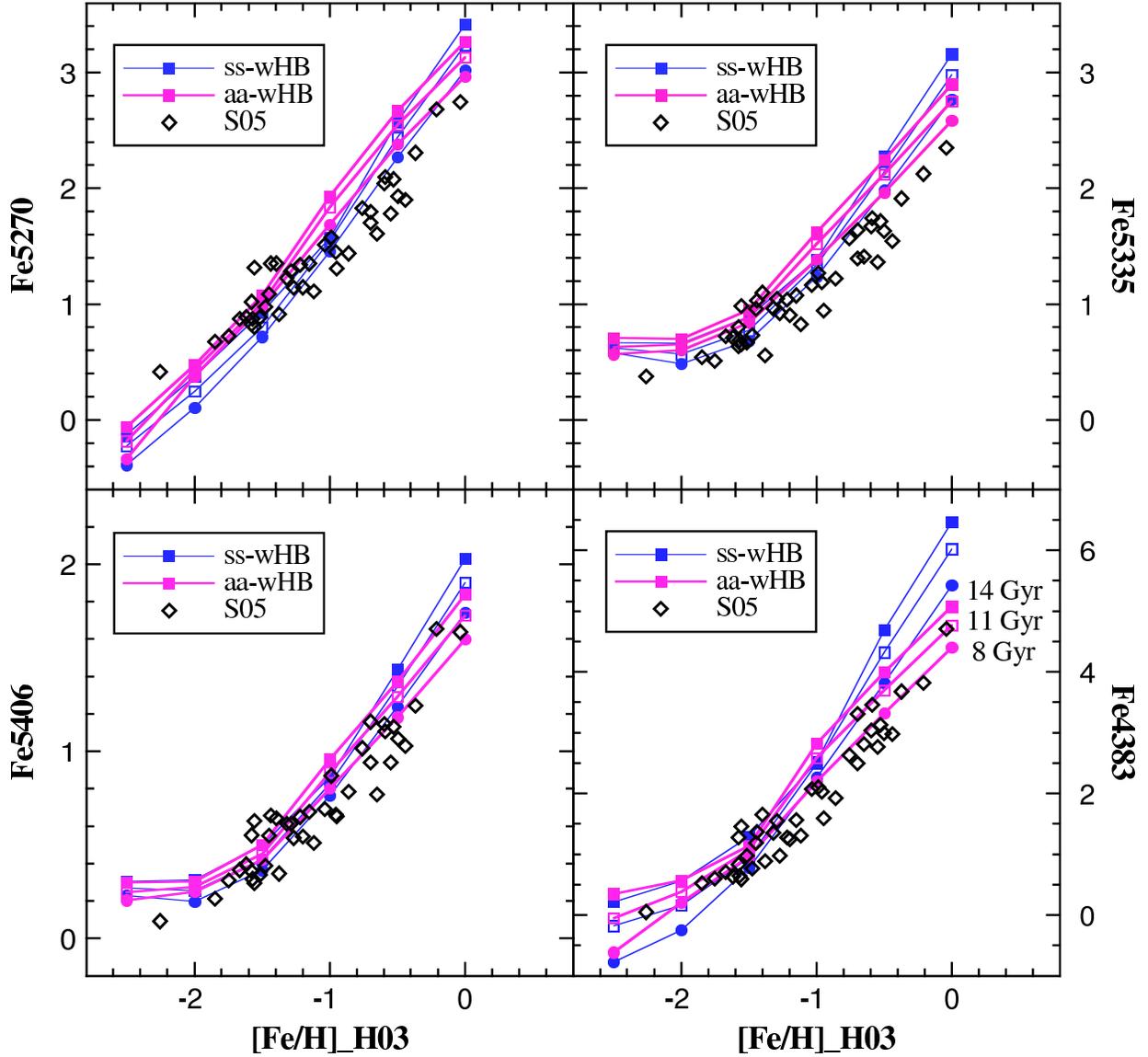}
\caption{Same as Figure 16, but the [Fe/H] values of MWGCs are from Harris 
2003 compilation instead of Table 1 of Schiavon et al. (2005).  Compared to 
Figure 16, the matches are not very favorable, especially 
at $-$1.0 $<$ [Fe/H] $<$ $-$0.5.  
\label{fig17}}
\end{figure}

\begin{figure}
\epsscale{.8}
\plotone{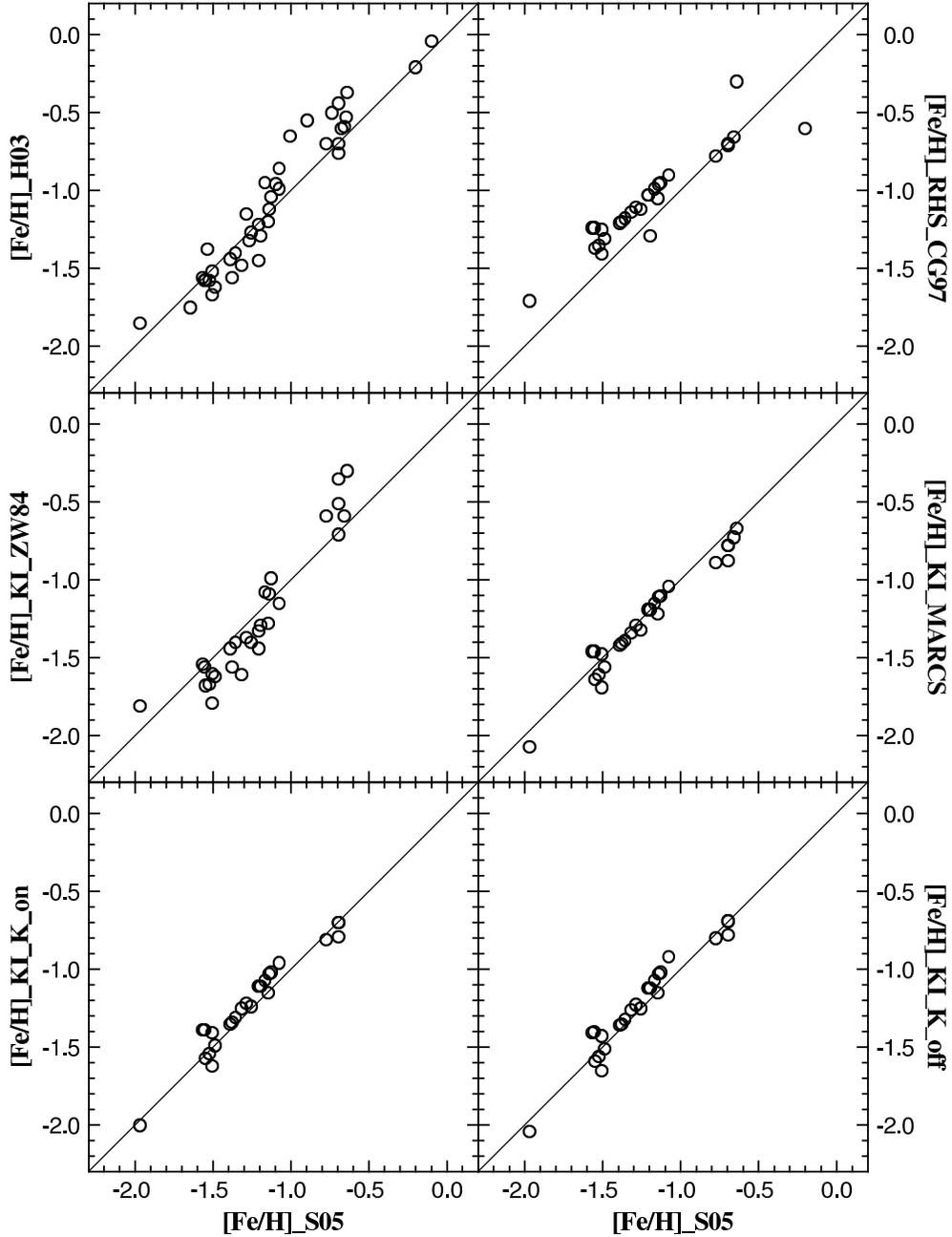}
\caption{Several different cluster [Fe/H] compilations from the literature are
compared with that of Table 1, Schiavon et al. (2005).  They are the Harris
2003 compilation (top left), Carretta \& Gratton (CG97) in Table 2 of
Rutledge, Hesser, \& Stetson (1997) (top right), and Zinn \& West (ZW84)
(middle left), MARCS (middle right), Kurucz-on (bottom left), Kurucz-off
(bottom right) in Table 7 of Kraft \& Ivans (2003), respectively (see text for
details).
\label{fig18}}
\end{figure}

\end{document}